%% file: main_ijhcs.tex
\documentclass[preprint,12pt]{elsarticle}

\usepackage{amssymb}
\usepackage{amsmath}
\usepackage{tabularx}
\usepackage{graphicx}
\usepackage{subcaption}
\usepackage{multirow}
\usepackage{enumitem}
\usepackage{booktabs}
\usepackage{xcolor}
\usepackage{colortbl}
\usepackage{xurl}
\newcommand{\up}{\raisebox{0.5pt}{\scalebox{0.8}{$\uparrow$}}}
\newcommand{\down}{\raisebox{0.5pt}{\scalebox{0.8}{$\downarrow$}}}

\newcommand{\sigone}[1]{\textcolor{blue!80!black}{#1}}
\newcommand{\sigtwo}[1]{\textcolor{red!60!black}{#1}}
\newcommand{\best}[1]{\textcolor{red!60!black}{\underline{#1}}}

\usepackage{pdfpages} % 关键宏包

\newcommand{\revise}[1]{\textcolor{black}{#1}}
\newcommand{\revtwo}[1]{\textcolor{black}{#1}} % Round-2 revision tracking
\newcommand{\newcite}[1]{\textcolor{black}{\citep{#1}}} % Newly added citations (from prior work), shown in RED

\journal{International Journal of Human-Computer Studies}

\begin{document}

\begin{frontmatter}



% \title{LLM Rationales as Auditable Trust-Calibration Interfaces: Effects on Trust, Decisions, and Visual Attention}
% \title{Auditable Trust Calibration with LLM Rationales: Effects on Trust, Decisions, and Visual Attention}
% \title{When LLM Rationales Become User-Facing: Effects of Auditable Trust Calibration with LLM Rationaleson Trust Calibration, Decisions, and Visual Attention}
% Auditable Trust Calibration: How User-Facing LLM Rationales Influence Trust, Decisions, and Visual Attention
\title{When LLM Rationales Become User-Facing: Effects on Trust Perception, Decision-Making, and Gaze Behaviors}

% ✅ Author contributions:
% Xin Sun: Conceptualization, study design, experimental/technical setup, data analysis, writing;
% Shu Wei: Conceptualization, data analysis, writing, revision;
% Ting Pan & Yajing Wang: Conceptualization, data collection;
% Jos Bosch: Funding support, Ethical application;
% Isao Echizen: Funding support;
% Abdallah El Ali: Conceptualization, revision;
% Saku Sugawara: Conceptualization, revision;

% \author[a,b]{Xin Sun\corref{cor1}}
\cortext[cor1]{Corresponding author}
% \ead{x.sun2@uva.nl}

\author[a,b]{Xin Sun}
\author[b]{Ting Pan}
\author[b]{Yajing Wang}
\author[c]{Shu Wei}
\author[b]{Jos A. Bosch}
\author[a,d]{Isao Echizen}
\author[e,f]{Abdallah El Ali\corref{cor1}}
\author[a,d]{Saku Sugawara\corref{cor1}}

\affiliation[a]{
  organization={National Institute of Informatics},
  country={Japan}
}

\affiliation[b]{
  organization={University of Amsterdam},
  country={The Netherlands}
}

\affiliation[c]{
  organization={Yale School of Medicine},
  country={United States}
}

\affiliation[d]{
  organization={University of Tokyo},
  country={Japan}
}

\affiliation[e]{
  organization={Centrum Wiskunde \& Informatica (CWI)},
  country={The Netherlands}
}

\affiliation[f]{
  organization={Utrecht University},
  country={The Netherlands}
}

% =====================================================================================

%% Abstract
\begin{abstract}
Large language models (LLMs) increasingly show step-by-step reasoning rationales alongside their answers, turning reasoning from an internal model capability into a user-facing interface feature.
Yet it is unclear whether such rationales help users judge when trust is warranted or merely persuade through fluent reasoning.
% We frame this as auditable trust calibration: rationales should help users judge whether an answer is supported by evidence, whether its certainty is justified, and whether to accept, question, or verify the advice.
We address this gap through the lens of auditable trust calibration: user-facing rationales should help people inspect whether an answer is warranted by evidence, whether expressed certainty is justified, and whether AI advice should be accepted, questioned, or independently verified.
We test this framing in factual verification through two linked studies.
Study~1, an online experiment (N=68), manipulated rationale presentation format (instant, delayed, on demand), rationale correctness (correct, incorrect), and certainty framing (none, certain, uncertain).
Study~2, a controlled eye-tracking study (N=54), examined how no-, correct-, and incorrect-rationale conditions were associated with users' trust, decision-making, and eye movement patterns.
% \revise{Study~1 showed no reliable presentation-format effects; instead, rationale correctness most consistently raised trust in the information, trust in the LLM system, and decision confidence, while certainty framing also shifted trust- and decision-related outcomes.}
\revise{Study~1 showed no reliable presentation-format effects; instead, rationale correctness and certainty framing influenced the trust in the information, trust in the LLM system, and decision confidence.}
\revise{In Study~2, incorrect rationales drew more attention to the supporting evidence and larger pupil diameter while the rationale was viewed, consistent with greater cognitive effort, though within-rationale fixation counts and durations did not differ by correctness.}
\revtwo{Incorrect rationales also lowered trust in the LLM system relative to showing no rationale, whereas the no-rationale difference was weaker for trust in information.}
A post-hoc predictive modeling analysis of gaze data from Study~2 further showed that gaze features carried predictive signal for trust- and decision-related user states.
This work challenges the assumption that more reasoning is always better and supports rationale designs that are selective, linked to evidence, calibrated in how they express certainty, and easier to verify.
% /, under reliable AI answers, support rationale designs that are selective, linked to evidence, calibrated in how they express certainty, and easier to verify.
\end{abstract}

\begin{keyword}
LLM rationales \sep Trust calibration \sep Eye tracking \sep LLM-assisted decision-making \sep Factual verification
\end{keyword}

\end{frontmatter}

% \begin{figure}[!ht]
%   \center
%   \includegraphics[width=0.996\textwidth]{figures/lab_setup_a.png}
%   \caption{The hardware setup for presenting the text stimulus and collecting physiological signals, eye movement, and pupil dilation.}
% \end{figure}



% ================================================================================

\input{main/1-Introduction}

\input{main/2-Related}

\input{main/3-Study1-methods}

\input{main/3-Study1-results}

\input{main/4-Study2-methods}
\input{main/4-Study2-results}

\input{main/5-Study3-methods}
\input{main/5-Study3-results}

\input{main/6-Discussion}

\input{main/7-Conclusion}

% ================================================================================

\section*{Ethical Statement}

This work involved human subjects and received ethical approval from the institute's ethics committee. 
All participants provided informed consent before the study. 
All data were collected only for research, anonymized and reported in aggregate.
Gaze-based user-state prediction can raise privacy and manipulation concerns. 
For analysis, raw gaze samples were converted into aggregate AOI-level features for prediction tasks.
Any future deployment of the proposed user-aware system should be further ethically reviewed and must be informed, transparent, opt-in, privacy-preserving, and designed to support user autonomy rather than persuasion, for example, in line with relevant transparency obligations, such as those in the EU AI Act~\citep{eu_act}.

Besides, LLMs' rationales and certainty framing may mislead users. To minimize risk, we used non-sensitive fact-verification tasks and did not collect personal or identifiable information. 
This work aims to support user-protective and human-centered AI by guiding how LLMs' rationales and certainty are presented in AI \revtwo{decision-support systems} to inform safer practices that encourage appropriate trust and reliance.

% We carefully considered the ethical implications of this research, including the risks of handling sensitive user information and potential biases in generated content. 
% This study adheres to the ethical standards of our institute, with strict measures to safeguard participant data and prevent harmful or biased AI outputs. 
% Informed consent was obtained from all participants, ensuring anonymity and the option to withdraw at any time without repercussions. 

% ================================================================================

\section{AI Usage Disclosure}
\label{sec:appendix_ai_disclosure}

We used OpenAI GPT-based tools (i.e., GPT-5.4) for language editing to improve clarity and conciseness. 
\revise{This writing assistance was separate from experimental-material preparation. The factual claims used in the studies were drawn from StrategyQA, and the final participant-facing answer and rationale texts were curated, edited, and checked by the authors as controlled experimental materials before deployment.}
All study design, analysis, literature review, and writing were conducted and verified by the authors.

\newenvironment{acks}{\section*{Acknowledgments}}{}
% \begin{acks}

% \end{acks}

\newpage

\bibliographystyle{elsarticle-num} 
\bibliography{main_ijhcs} %% consolidated bib: Section A = currently-cited refs; Section B = newly added (red) refs




% %% For numbered reference style

%   \textit{\LaTeX: a document preparation system},
%   Addison Wesley, Massachusetts,

% \end{thebibliography}

% \clearpage
% \newpage

% \appendix
% \section*{Supplementary Material}
% \label{supplement}

% \includepdf[pages=-]{figures/IJHCS_Supplement-Stimuli.pdf} % `pages=-` 表示插入所有页

\end{document}

%% file: main/1-Introduction.tex
\section{Introduction}

Large language models (LLMs) increasingly present step-by-step reasoning rationales alongside their answers. In model-centric research, such reasoning is studied mainly as a technique for improving performance on tasks that require decomposition or multi-step inference~\citep{cot,lrm_survey}. In interactive systems, however, the same reasoning text becomes a visible part of the user interface: users receive not just an answer but a narrative account of how it was produced, which they may read as evidence, justification, a competence signal, or an invitation to verify~\citep{users_perceive_cot}. This makes LLM reasoning a component in human-computer interaction, rather than only a model-capability problem: once exposed, reasoning can alter users' attention, trust, confidence, and willingness to act on LLM's advice.

This shift matters because more trust is not inherently better in AI-assisted decision-making. A long tradition in automation and human-AI interaction argues that the goal is \emph{appropriate trust}: relying on a system when it is competent for the task and withholding trust when it is unreliable, uncertain, or out of scope~\citep{measure_trust_automation,Wischnewski}. Calibration is therefore relational and task-sensitive: a user may trust a system in general yet doubt a specific answer, or accept an answer while distrusting its reasoning. LLM rationales complicate this because they provide a persuasive natural-language account at the very moment users must decide whether to rely.

The central risk is miscalibration. Natural-language rationales can be fluent, coherent, and specific even when they are logically flawed, factually incorrect, inconsistent with the evidence, or not faithful to the model's actual decision process~\citep{reasoning_incorrect_1,reasoning_incorrect_2,inconsistency_llm_reasoning}. Such rationales may create a sense that an answer has been carefully derived even when the reasoning merely rationalizes an answer after the fact. They may also contain verbal certainty cues, such as statements that the model is confident or unsure, which can influence users independently of the answer's actual correctness or the rationale's epistemic quality~\citep{llm_uncertainty,measure_confidence,kim2024uncertainty}. 
In factual verification and other decision-support settings, these cues may raise subjective confidence without improving accuracy, or trigger distrust when reasoning errors become visible. Recent work on LLM-assisted fact-checking and question answering shows the same tension: explanations can help users verify information but can mislead them when the explanation is wrong~\citep{si2024verify,kim2025reliance}. Thus, LLM rationales are not neutral add-ons. 
They are interface-level elements that can support trust when they help users verify an answer, but can miscalibrate trust when they substitute fluency or certainty for the warranted evidence.

Prior work in explainable AI (XAI) and AI-assisted decision-making has established that explanations affect trust, confidence, and mental models~\citep{xai,confidence_effect_in_decision_1,confidence_effect_in_decision_2,impact_xai_decision_making}. This literature also shows a more troubling pattern: AI explanations do not consistently improve decision quality. They can encourage over-trust when users accept AI advice without engaging critically with the task itself~\citep{explanation_ai_overreliance,bansal2021,schemmer2023}. However, much of this work has examined compact or bounded explanations, such as feature attributions, saliency maps, or short decision rules. LLM rationales differ because they are open-ended, natural-language, and often conversationally persuasive. These properties make the general question ``do explanations help?'' too blunt for user-facing LLM rationales. The more precise question is how users inspect, interpret, and rely on rationales whose correctness, certainty, and presentation can vary.

Despite these risks, the mechanisms by which user-facing LLM rationales influence trust-related judgments and may support or undermine future trust calibration remain under-specified. Prior work has examined whether reasoning improves model performance~\citep{cot}, how users perceive LLM explanations~\citep{users_perceive_cot}, and how uncertainty communication influences AI-assisted decisions~\citep{bhatt2021,bussone2015}. 
What remains unclear is how rationale properties such as correctness, certainty framing, and presentation format affect trust and advice adoption, and how users behaviorally inspect rationales while making decisions. 
Without this evidence, we cannot determine when showing reasoning supports appropriate trust and when it risks miscalibration.

We study this gap through the lens of \emph{auditable trust calibration}. \revtwo{We use this term as a design goal, not as a directly measured outcome.} Ideally, user-facing rationales should help users decide whether an answer is supported by evidence, whether the model's certainty is warranted, and whether AI advice should be accepted, questioned, or verified. In the present work, we examine how rationale quality changes users' trust judgments, advice adoption under reliable AI answers, without measuring calibration accuracy itself.
This perspective is grounded in prior work on explanation engagement and trust~\citep{impact_xai_decision_making,explanation_ai_overreliance,responsible_ai}, in which user-facing rationales are evaluated by whether they help users rely on AI advice appropriately, highlighting three roles of rationales. 
First, rationales can serve as \emph{calibration cues}: they may shift trust by signaling competence, uncertainty or inconsistency. Second, rationales can function as \emph{audit objects}: when they are inspectable, users can compare the reasoning with the answer and the available evidence to judge whether the AI's advice is well supported. Third, rationales create \emph{interaction costs}: they support verification only when users are willing and able to spend attention reading, comparing, and questioning them. 
From this view, a rationale is useful not because it is fluent, but because it makes the relationship among answer, evidence, uncertainty, and user action easier to inspect.

We investigate this framing through two connected studies and a post-hoc predictive modeling analysis (Fig.~\ref{fig:procedure}).
Factual verification is a useful context because it poses a clear reliability problem that users must judge whether a binary answer and its rationale are correct, and it separates answer acceptance from rationale inspection: a user may adopt an answer on a credible rationale, reject it when the rationale exposes a flaw, or turn to external evidence when the rationale creates doubt.
Guided by this framing, we ask three research questions.

\begin{enumerate}[leftmargin=*,label=\textbf{RQ\arabic*.}]
    \item How do the rationale presentation format, rationale correctness, and certainty framing influence users' trust in the information, trust in the LLM system, decision confidence, and advice adoption?
\end{enumerate}

To answer RQ1, Study~1 uses an online experiment to probe the design space of user-facing rationales. It manipulates three properties central to rationales: how the rationale is revealed (instant, delayed, or on demand), whether the rationale is correct or incorrect, and whether the rationale is accompanied by no certainty cue, a certainty cue, or an uncertainty cue.
Together, these manipulations allow us to test whether users are influenced more by how rationales are presented or by whether the reasoning is reliable and how confidently it is framed.
\revise{Study~1 shows \revtwo{no reliable presentation-format effects}. Instead, rationale correctness had the clearest effects across outcomes, increasing trust in the information, trust in the LLM system, decision confidence, and advice adoption, while certainty framing also shifted trust-, confidence-, and adoption-related judgments.}

However, self-reported trust and decisions do not reveal \emph{how} users process a rationale. A user can report trust after a trial without revealing whether they read the rationale carefully, relied mainly on the AI answer, or moved between these sources of information. 
This creates a second gap: we need \revtwo{process evidence} for how rationales are visually inspected during participants' decision-making. We therefore ask:

\begin{enumerate}[leftmargin=*,label=\textbf{RQ\arabic*.},resume]
    \item How do the presence and the correctness of the LLM's rationales change users' trust, decision-making, and how they visually inspect the answer, evidence, and rationale?
\end{enumerate}

Study~2 addresses RQ2 by examining the behavioral mechanism behind rationale inspection in a controlled lab eye-tracking study (N=54).
It compares no-, correct-, and incorrect-rationale conditions while measuring the same self-reports as Study~1, with additional eye movement across four Areas of Interest (AOIs): context (factual question and evidence), LLM's answer, LLM's rationale, and self-reported rating.
Eye tracking is used as the process evidence of attention allocation, not as a direct measure of trust~\citep{Kohn_Spencer,eyetracking_methods}.
The results show that rationale correctness influenced inspection patterns in specific regions: \revise{incorrect rationales were associated with more scanning of the supporting evidence and larger pupil diameter while viewing the rationale, whereas fixation frequency and duration within the rationale region itself did not reliably differ by correctness}. 
Study~2 therefore connects the outcome effects observed in Study~1 to the inspection process itself.

Ultimately, if gaze reveals how users inspect rationales, it may carry predictive signals about user states for both trust and decision-making.
This matters because self-reports are collected after the decision and may miss the process through which that decision was formed. We therefore ask:

\begin{figure*}[!ht]
    \centering
    \begin{subfigure}{0.999\textwidth}
        \centering
        \includegraphics[width=\linewidth]{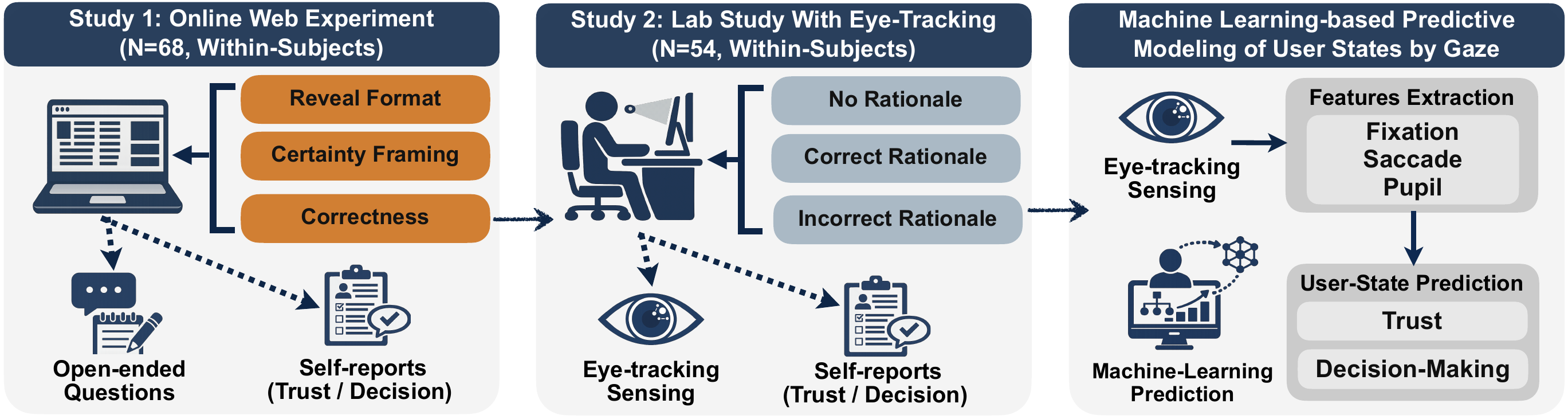}
        \caption{Overview of the entire investigation with three steps.}
    \end{subfigure}

    \vspace{2.6mm}

    \begin{subfigure}{0.999\textwidth}
        \centering
        \includegraphics[width=\linewidth]{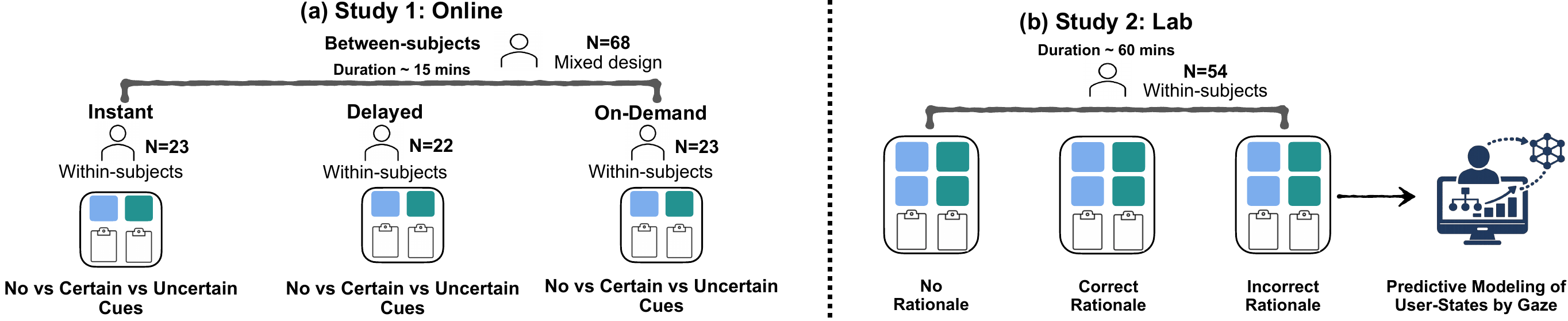}
        \caption{Procedure of two studies. Study~1 is an online study, and Study~2 is a lab eye-tracking study.}
    \end{subfigure}

    \vspace{-0.0mm}
    \caption{
    Overview of the three-part investigation.
    \textbf{\emph{Study 1}} maps the design space of user-facing LLM rationales by manipulating presentation format, rationale correctness, and certainty framing in an online study.
    \textbf{\emph{Study 2}} examines visual inspection behavior in a lab eye-tracking study comparing no-rationale, correct-rationale, and incorrect-rationale conditions.
    \textbf{\emph{Predictive modeling}} uses gaze features from Study~2 to explore whether visual behavior carries retrospective, task-specific information about trust- and decision-related user states.
    (a) Conceptual overview of the three steps. (b) Study procedures.
    }
    \label{fig:procedure}
    \vspace{-2.0mm}
\end{figure*}

\begin{enumerate}[leftmargin=*,label=\textbf{RQ\arabic*.},resume]
\item To what extent can gaze features predict trust- and decision-related user self-reported states in LLM-assisted factual verification?
\end{enumerate}

To answer RQ3, we conduct \revise{a secondary, exploratory predictive modeling analysis using gaze features from Study~2, to classify trust- and decision-related user self-reported states in LLM-assisted factual verification tasks.} 
We do not treat gaze as a direct reading of trust. Instead, we use predictive modeling to test whether behavioral traces contain useful information about the user states that matter for trust calibration.
This work makes four contributions. 
First, \revise{we contribute an auditable trust calibration design lens as a normative framework for evaluating user-facing LLM rationales.}
Second, we map the design space of user-facing rationales, isolating how correctness, certainty framing, and presentation format affect trust and advice adoption. 
Third, using eye tracking, we provide process-level evidence for how users inspect the rationale, answer, and evidence while deciding. 
Fourth, \revise{we conduct an exploratory analysis of whether gaze features carry retrospective, task-specific signals about trust- and decision-related outcomes}. 
Together, these contributions challenge the assumption that showing more reasoning is always better and instead support rationale designs that are selective, careful in how they express uncertainty, and easier to verify.

%% file: main/2-Related.tex
\section{Related Work}

\subsection{Trust Calibration and Reliance in AI-Assisted Decision-Making}

Trust is central to AI-assisted decision-making, but the goal is not simply to make users trust AI more.
In decision-support settings, high trust can be harmful when the system is wrong, and low trust can be harmful when the system is useful. Prior work in automation and human-AI interaction therefore emphasizes \emph{appropriate reliance}: users should rely on AI when it is likely to help and should withhold reliance when the system is uncertain, unreliable, or outside its competence~\citep{measure_trust_automation,Wischnewski,trust_calib_hri,automation_bias_review}. This distinction is important for our work because LLM rationales may increase trust without necessarily improving the quality of users' decisions.

Trust and reliance are related, but they are not the same construct. 
In classic trust theory, trust is defined by Mayer et al. as a willingness to be vulnerable to another actor under uncertainty, based on perceived trustworthiness attributes such as ability, benevolence, and integrity~\citep{organizational_trust_model}. 
In automation and human-AI interaction research, this idea is adapted to technical systems: trust reflects users' expectations about a system's competence, reliability, predictability, or helpfulness in a given context, whereas reliance refers to the behavioral act of accepting, following, or acting on system advice~\citep{trust_automation,measure_trust_automation,Wischnewski,hoff_bashir2015,Vereschak2024,trust_ca}. 
This distinction matters because trust does not always translate into reliance, and reliance does not always imply well-calibrated trust. Users may trust an AI system in general but reject a specific recommendation, follow a recommendation while remaining unsure about the reasoning behind it, or become confident in their own decision after using AI advice without increasing system-level trust. 
For user-facing LLM rationales, this distinction is especially relevant because a fluent rationale may increase perceived ability or competence without necessarily improving the correctness of the answer or the appropriateness of reliance. 
In this work, we therefore distinguish trust and decision-related measures. 
This allows us to examine whether rationales merely change users' trust level toward the AI or whether they also affect concrete decision behaviors.

Explanations are often proposed as a way to improve this calibration. Explainable AI (XAI) research has shown that explanations can shape users' understanding, confidence, trust, reliance, and mental models~\citep{xai,confidence_effect_in_decision_1,confidence_effect_in_decision_2,impact_xai_decision_making,accuracy_fidelity_trust,explanation_trust_recsys,Wang2023-nm,contradictory_explanations}. 
However, the explanations can make an answer feel more understandable or credible without helping users detect errors. 
They can also increase reliance on incorrect advice when users accept the explanation as sufficient justification~\citep{explanation_ai_overreliance,bansal2021,schemmer2023,imperfect_xai,misleading_explanations}. In this sense, explanation can support calibration, but can also create overreliance.
Prior work argues that explanation use depends on this cost-benefit tradeoff: users are more likely to benefit from explanations when it makes verification easy and when explanation is clearly connected to the decision evidence~\citep{explanation_ai_overreliance}. If an explanation is difficult to inspect, disconnected from evidence, or persuasive without being reliable, it may increase users' confidence without improving their decisions.

This literature gives our work a theoretical starting point.
We do not evaluate LLM rationales by asking only whether they influence trust. 
We also ask whether they help users \revise{rely appropriately}: whether users can use the rationale to judge when an AI answer is warranted, when it should be questioned, and when external evidence should be checked. This is especially important for LLM rationales because they are not simple feature weights or short rules. 
They are natural-language reasoning traces that can sound like careful deliberation. 
As a result, they may function not only as explanations, but also as confidence cues, social signals, and evidence-like narratives.

% \subsection{User-Facing LLM Rationales: Transparency Limits, Presentation, and Certainty Framing}
\subsection{User-Facing LLM Rationales: Development, Design and Challenges}

Research on large reasoning models or reasoning-based prompting techniques, such as chain-of-thought~\cite{cot}, has mainly examined whether step-by-step reasoning improves model performance~\citep{cot,lrm_survey}. This model-centered work is important, but it does not answer a user-centered question: what happens when the reasoning trace is shown to people? Once a rationale appears in an interface, it becomes more than a computational artifact. It becomes part of the user's evidence environment. Users may read it as an explanation, a justification, a signal of competence, or a reason to accept the model's answer~\citep{users_perceive_cot}. This gives user-facing LLM rationales a different role from many traditional XAI outputs, such as feature attributions, saliency maps, or compact decision rules~\citep{xai_shape_lime,xai_shape}.

A key challenge is that LLM rationales are not guaranteed to be faithful or correct. Prior work shows that chain-of-thought explanations can be plausible but unfaithful, can rationalize biased or incorrect answers, and can be inconsistent with the final response~\citep{reasoning_incorrect_1,reasoning_incorrect_2,inconsistency_llm_reasoning,reasoning_incorrect_3}. This creates a transparency problem. A rationale may look like an audit trail, but it may instead be a fluent post-hoc narrative. Users may treat the rationale as evidence that the answer has been carefully derived, even when the reasoning does not accurately support the answer or reflect the model's generative process.

For this reason, rationale correctness is not only a model-quality issue, it is also an interface issue. A correct rationale can help users connect an answer to the relevant evidence. An incorrect rationale can expose a flaw, create confusion, or mislead users if it remains fluent and convincing. In factual verification tasks, this matters because users must decide whether the answer is supported by the evidence, not merely whether the rationale sounds reasonable. 
Correctness therefore determines whether rationales can function as a useful audit object or becomes another source of miscalibration. Recent studies on LLM-assisted question answering and fact-checking report the same pattern: explanations can improve verification efficiency while still increasing overreliance when explanations are wrong, and sources or inconsistency cues can help restore more appropriate reliance~\citep{si2024verify,kim2025reliance,yao2025factcheckingaigeneratednewsreports,Hallucinations}.

How the rationale is presented may also matter. Human-AI collaboration is shaped not only by what information is shown, but also by when and how users encounter it~\citep{AI_assisted_decision_making,progressive_disclosure,fernandes2026explainingmuchunderstandinglarge}. 
A rationale can be shown immediately, delayed until after an initial answer, or made available on demand. Immediate presentation lowers access cost but may encourage passive reading. Delayed presentation may encourage users to form an initial judgment before seeing the reasoning. On-demand presentation gives users control, but also requires them to actively request the rationale. These formats change the interaction cost of inspecting reasoning, but it remains unclear whether they matter as much as the reliability of the rationale itself.

Certainty framing adds another important design dimension. 
LLMs often express confidence or uncertainty. Prior work shows that such cues can influence users' trust, satisfaction, confidence, and decisions~\citep{llm_uncertainty,measure_confidence,kim2024uncertainty}. 
A confident rationale may appear more credible even when the reasoning is flawed, whereas an uncertainty cue may reduce trust even when the answer is correct. 
Certainty framing therefore acts as an epistemic cue: it tells users how strongly the system appears to stand behind its reasoning. 
The risk is that this cue may not be calibrated to the actual quality of the rationale.

Together, this work motivates our exploration. Prior research has examined reasoning, explanation presentation, uncertainty communication, and explanation quality, but these factors are often studied separately. We therefore test them together in the context of user-facing LLM rationales. 
% Study~1 asks whether users are more influenced by the mechanics of rationale disclosure or by the rationale's epistemic quality: whether it is correct and how certainty is expressed.

\subsection{Sensing and User-State Modeling in Human-AI Interaction}

Self-report measures are necessary for understanding people's perceptions, but they cannot show how users process AI generations during interaction~\citep{sun_trust}. 
A user may report high trust after a task, but that report alone does not show whether the user carefully read the rationale, checked it against evidence, ignored the evidence, or accepted the answer after a quick scan. 
This limitation is central to our auditability framing. If rationales are meant to support verification, we need evidence about whether users actually look at the rationale and the contextual information needed to evaluate it.

Eye tracking can provide this kind of process evidence. Fixation, saccade, and pupil features are commonly used to study visual attention and information processing~\citep{eyetracking_methods,sun_trust,eyetracking_survey,eyetracking_cognitive_2,eye_movements_decision,pupil_cognitive_load,Wang2001-hf}. 
In human-AI interaction, gaze and physiological signals have been used to study cognitive load, uncertainty, information seeking, credibility judgments, and trust-related states~\citep{eyetracking_cognitive_1,physio_search,Abdrabou_Yasmeen,Ajenaghughrure_modeling,trust_predict_hri,sun_trust,Chiossi2024,SUMER2021106909,wearable_sensing}. 
For LLM rationales, gaze is useful because it can show whether users attend mainly to the answer, inspect the rationale, check external evidence, or move between these regions.

Despite this, gaze must be interpreted carefully. It is not a direct measure of trust, nor does it reveal a user's mental state by itself~\citep{Cacioppo}. 
The same gaze pattern can mean different things depending on the contexts. 
% Longer fixation on a rationale may reflect careful checking, confusion, persuasion, or difficulty. More movement between evidence and rationale may reflect verification effort, uncertainty, or conflict monitoring. 
The value of gaze is therefore that it gives behavioral evidence about how users process information while trust and decisions are being formed.

There is still limited evidence on how gaze changes when users interact with correct versus incorrect LLM rationales. This is an important gap because rationale correctness should affect not only trust, but also how users inspect the information. 
If a rationale is correct and coherent, users may integrate it with the answer. If a rationale is inconsistent with the answer, users may shift attention toward the evidence or show patterns consistent with greater cognitive effort. 
Our investigation addresses this gap by measuring how users visually inspect evidence, LLM answers, rationales, and \revtwo{response/rating} areas under no-, correct-, and incorrect-rationale conditions.

The same gaze features may also contain information relevant to user-state modeling. 
Prior work suggests that sensed signals can support prediction of task-related states such as cognitive load and trust, but such models depend strongly on the task, features, and target construct~\citep{Ajenaghughrure_modeling,trust_predict_hri,sun_trust,Akash_Kumar,Ahmad_Muneeb,Parikh2018EyeGF}. 
We therefore treat predictive modeling as a post-hoc analysis.
In this paper, gaze is used in two cautious ways: as process evidence for how users inspect rationales in Study~2, and as a test of whether visual behavior carries task-specific signal about trust- and decision-related states.

%% file: main/3-Study1-methods.tex
% =================================================================

\section{Study 1: Online Survey Study}

Study 1 was designed to map the design space of user-facing LLM rationales. 
Once reasoning rationales are exposed to users, they become interface elements with several controllable properties: first, they can be revealed immediately, delayed, or on demand; 
second, they can contain correct or incorrect reasoning, and lastly, they can be framed with certainty or uncertainty. 
These choices may affect users in different ways. 
% Presentation format may change how easily users access the rationale, correctness determines whether the rationale is a reliable audit trace, and certainty framing signals how strongly the system appears to endorse its own reasoning. 
The objective of Study 1 was therefore to examine which of these properties most strongly influences users' trust, decision confidence, and willingness to follow LLM advice.

\input{tables/stimuli_examples}

\subsection{Methods}

\subsubsection{Study Design and Procedure}

We ran an online experiment with a mixed design: 2 (Rationale Correctness: correct vs.\ incorrect; within-subjects) $\times$ 3 (Certainty Cue: none vs.\ uncertain vs.\ certain; within-subjects) $\times$ 3 (Rationale Presentation: instant vs.\ delayed vs.\ on-demand; between-subjects). This design separated two audit-relevant properties of the rationale itself (whether the reasoning was correct and how certain it sounded) from the presentation format through which the rationale was made available.

\begin{figure*}[!ht]
    \centering
    \includegraphics[width=0.99\linewidth]{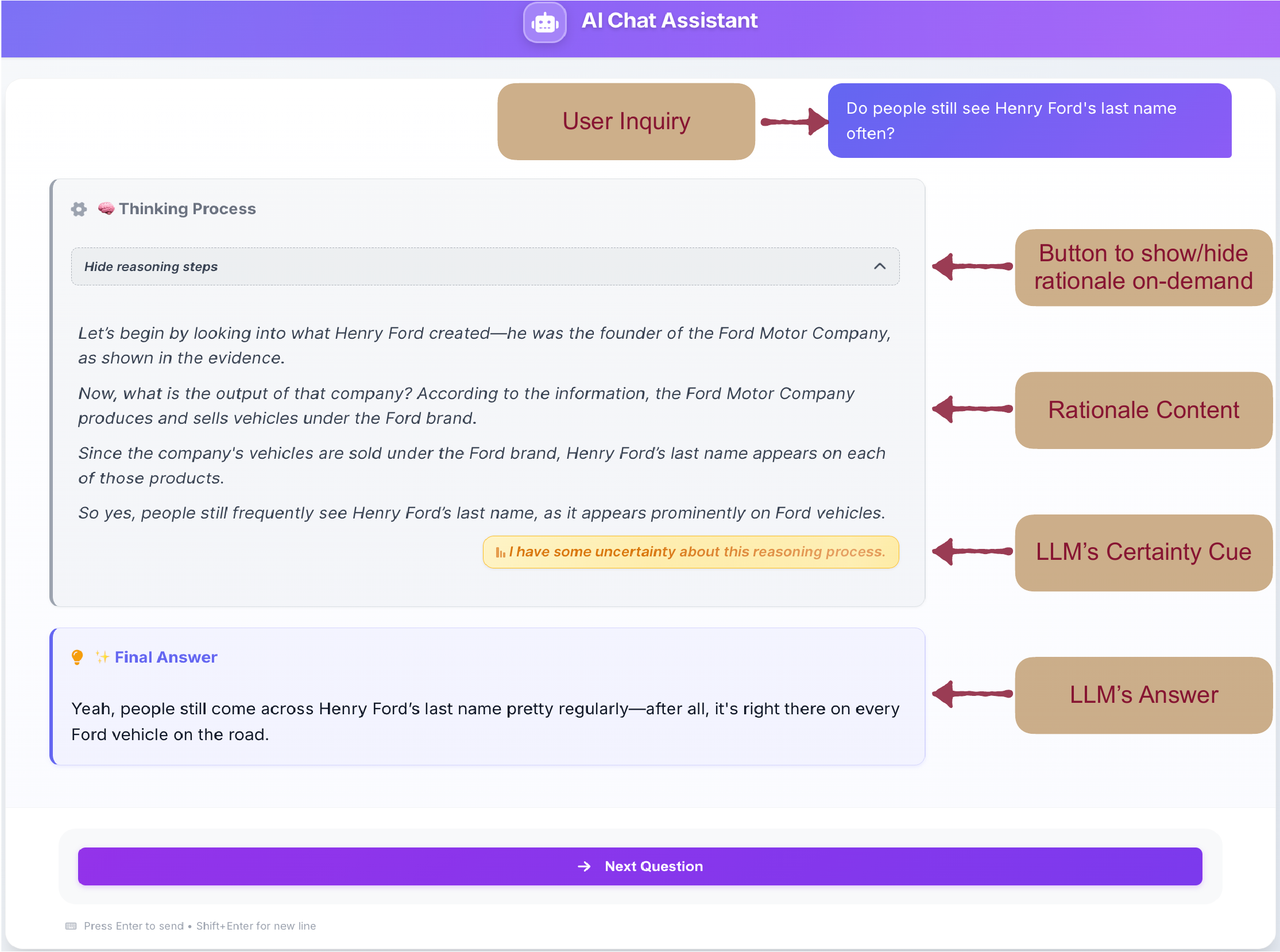}
    \vspace{-1.0mm}
    \caption{Study 1 online interface example. The screenshot illustrates the on-demand rationale interface, in which participants could reveal or hide the rationale. Across presentation-format conditions, participants were given a factual verification query; they then viewed the LLM answer with the assigned rationale presentation format, and reported their trust level, decision confidence, and advice adoption choose.}
    \label{fig:interface}
    \vspace{-2.0mm}
\end{figure*}

\textbf{Rationale presentation.} In the \textit{instant} condition, the answer and rationale appeared together. In the \textit{delayed} condition, the answer appeared first and the same rationale appeared after a brief, length-proportional delay, simulating a short generation interval while keeping the content unchanged following~\cite{response_delay}. 
In the \textit{on-demand} condition, the rationale was hidden by default and participants could reveal or hide it via a \emph{Show/Hide thinking steps} button (shown in Fig.~\ref{fig:interface}). 
Across the three presentation conditions, the factual query, LLM answer, rationale text, certainty cue, response options, and questionnaire items were kept constant; only the timing or user control of rationale visibility varied.
In the quantitative analyses, presentation format was therefore treated as an assigned interface factor rather than as a measure of how long each participant actually viewed the rationale.

% \revise{
% \textbf{Certainty cues.} The certainty manipulation was implemented as a brief cue appended to the rationale (one sentence). In the \textit{certain} condition, the cue expressed high confidence (e.g., ``I am confident in my reasoning.''), whereas in the \textit{uncertain} condition it expressed low confidence (e.g., ``I am not fully sure about my reasoning.''). The \textit{none} condition included no such cue. We used an explicit cue to isolate the causal effect of certainty framing independently of other stylistic signals (e.g., verbosity or hedging embedded throughout the rationale).}
\textbf{Certainty cues.} We manipulated the model's certainty by adding a one-sentence cue at the end of the rationale (Fig.~\ref{fig:interface}). In the \textit{certain} condition, the cue conveyed high certainty (e.g., ``I feel very certain about how I think about this question''); in the \textit{uncertain} condition, it conveyed hesitation (e.g., ``I'm somewhat hesitant about how I think about this question''); and the \textit{no-certainty} condition does not reveal any cue. 

\textbf{Rationale correctness.} As demonstrated in Table~\ref{stimuli_example}, we manipulated correctness at the level of the reasoning rationale. In \textit{correct} trials, the rationale provided reasoning consistently aligned with the LLM's answer. In \textit{incorrect} trials, the factual query, binary answer, certainty cue, and interface layout were kept unchanged, but the rationale contained a controlled \revtwo{flaw} that made the reasoning trace inconsistent with the LLM answer.
Thus, correctness captured whether the displayed rationale was a consistent trace for the answer.
Across all conditions, the LLM's binary answer was always factually correct; only the reasoning trace varied. 
This means that advice adoption in this study reflects whether participants agreed with a correct answer, not whether they discriminated between correct and incorrect AI advice. 
The design therefore tests how rationale quality influences trust and adoption when the answer itself is reliable, rather than measuring overreliance on incorrect advice (see Limitations).

The study procedure is shown in Fig.~\ref{fig:procedure}.
Each participant completed six trials, one for each combination of Rationale Correctness and Certainty Cue. 
Participants were randomly assigned to one of the three presentation-format conditions, making presentation format a between-subjects factor.
The six within-subject trials were counterbalanced in order. 
Participants first provided consent and completed a pre-survey. 
In each trial, they read a given factual verification query and the LLM's answer, inspected the rationale according to the assigned presentation format, and then completed the trial-level questionnaire. 
A final post-survey collected two open-ended responses about how the rationales affected their decision-making and trust ratings. 

% =================================================================

% =================================================================

% \vspace{0.3mm}
\subsubsection{Materials}
\label{study1_materials}

% =================================================================

\paragraph{\textbf{Interfaces and Stimuli}}
% % We developed a web (Fig.~\ref{fig:interface}) to allow participants ask a given factual claim simulating real conversational search. They saw LLM's answer with its reasoning rationale and certainty cue according to condition. 
% % Participants provided their decision and trust ratings after each trial.
% The factual claims are from a public fact-checking dataset~\citep{strategyqa} with model-generated reasoning steps. 
% For each claim, we created rationales: a correct rationale from dataset aligned with answer, and an incorrect rationale containing a controlled logical or factual flaw to the answer while preserving length and structure. Examples are shown in Append~\ref{sec:appendix_stimuli_examples}.

We developed a custom web interface (Fig.~\ref{fig:interface}) that simulates conversational information seeking. 
In each trial, participants entered a given factual query and then viewed the LLM's binary answer with its reasoning rationale and certainty cue (when applicable) according to the condition. The answer, rationale, and response items were shown in fixed interface regions so that only the manipulated rationale properties and presentation format varied across conditions.

The factual verification claims were adopted from a public fact-checking dataset ``StrategyQA''~\citep{strategyqa} and stepwise reasoning rationales were prepared for the experiment. 
We selected claims to be comparable in format and difficulty (short claims requiring evidence-based verification) and to avoid niche topics that would strongly depend on specialized knowledge. 
Trial presentation order was counterbalanced across participants. 
% The six selected claims were each assigned to a fixed within-subject condition cell\revtwo{; that is, each claim always appeared in the same Correctness $\times$ Certainty combination for every participant}. 
% \revtwo{Because each condition cell contained a single claim, condition effects are confounded with item-specific differences in content and difficulty. The reported condition effects should therefore be read as effects within this particular set of items rather than as item-generalized estimates (see Limitations).}

% \revise{The participant-facing answer and rationale texts were prepared as controlled experimental materials rather than inserted verbatim from the source dataset. Using each StrategyQA claim as the anchor, the authors iteratively drafted and edited matched correct and incorrect rationale versions so that the manipulation targeted the validity of the reasoning trace while keeping the final answer, interface layout, and overall wording style comparable across conditions. Before deployment, the authors checked all items for answer correctness and whether the incorrect version preserved the same final answer while introducing a deliberate flaw in the reasoning trace.}

% Across the two versions of each claim, we kept the length and structure of the rationale comparable, so that the correctness manipulation isolated the validity of the reasoning trace. Examples are shown in Table~\ref{stimuli_example}.

\revise{The participant-facing answer and rationale were prepared as controlled experimental materials rather than copied verbatim from the dataset. Each factual claim served as an anchor for matched correct and incorrect rationales, which were iteratively edited to keep the final answer, interface layout, rationale length, structure, and wording style comparable while varying the validity of the reasoning trace. 
Before deployment, all items were checked by authors to ensure that the final answer was factually correct and that the incorrect rationale preserved that answer while introducing a deliberate reasoning flaw. 
An example stimulus used for study is shown in Table~\ref{stimuli_example}.}

\input{tables/survey_demographics}

% ===============================================================================

\paragraph{\textbf{Measures}}
% Participants completed the measures after each trial:
After each trial, participants completed the following measures.

% \textbf{Decision:} accept vs.\ reject LLM’s binary answer~\cite{measure_trust}.
\noindent (1) \textbf{Decision (LLM's advice adoption):} whether participants \emph{accept or reject the LLM’s answer} (binary option). 
% \revise{In the quantitative models, advice adoption was coded as 1 when the participant's decision matched the displayed LLM answer and 0 otherwise.}

\noindent (2) \textbf{Decision confidence:} 7-point Likert scale~\cite{confidence_rating} (i.e., example item is ``How confident are you with the decision you made?'').

\noindent (3) \textbf{Trust in information:} participant's perceived trust in each LLM's response, 5-point Likert scale~\cite{trust_info_1,trust_info_2} (e.g., ``Is this information trustworthy?'').
% This item captures perceived credibility of the \emph{specific answer/content} for the current claim (i.e., ``Is this information trustworthy?'').

\noindent (4) \textbf{Trust in LLM system:} participant's perceived trust in LLM system using 5-point Likert scale~\cite{measure_trust_system}.
This item captures perceived trust of the \emph{system as a whole} across tasks (i.e., willingness to rely on the LLM in general), beyond the trust of any single response.

\noindent (5) \textbf{Manipulation check:} Participants indicated whether they perceived LLM's rationale as consistent with the LLM's answer or not. This item was used to verify that the correctness manipulation was perceived as intended.

\noindent (6) \textbf{Open-ended responses:} After all trials, participants answered two open-ended questions about how seeing the AI's ``thinking process'' affected their trust and decision-making, and which rationale features they found helpful or would improve. 
These supported the qualitative thematic analysis.
% and were not included as trial-level quantitative outcomes.

% =================================================================

% \vspace{-0.6mm}
\subsubsection{Participants and Data Analysis}

A priori G*Power~\cite{gpower} analysis suggested 48 participants to detect a medium effect ($f=0.25$) with 90\% power. 
We recruited 68 English-speaking participants via the crowd-sourcing platform Prolific~\cite{Prolific}. 
The sample was gender-balanced (33 female, 35 male). Most participants were 25-44 years old (70.6\%), all reported at least a bachelor's degree, and 60.3\% reported being very or extremely familiar with AI.
The study was approved by the institutional ethical committee. 
Participants were compensated according to platform standards. 
Demographic details are reported in Table~\ref{table:online_demographics}.

% ===============================================================================

\revise{For quantitative analysis, we aggregated trial-level responses to means for each factor level. Continuous outcomes included trust in information, trust in the LLM system, and decision confidence. Advice adoption was coded as whether the participant's Yes/No decision.}
\revise{Then the factor-wise analyses corresponding to the three experimental manipulations were conducted. 
Certainty Cue was tested with one-way repeated-measures ANOVAs~\cite{anova}. 
Rationale Correctness, a two-level within-subject factor, was also tested with a repeated-measures ANOVA; because this test is mathematically equivalent to the planned correct-versus-incorrect paired contrast (\(F=t^2\)), we interpret it as that contrast and report Cohen's \(d_z\) as the effect size. 
Rationale Presentation Format, a between-subjects factor, was tested with one-way between-subjects ANOVAs on participant-level means. 
Pairwise follow-up comparisons used paired \(t\)-tests~\newcite{paired_t_test} for within-subject contrasts and Welch \(t\)-tests for between-subject presentation-format contrasts, with Benjamini--Hochberg FDR correction applied within each outcome family. 
Effect sizes are reported as Cohen's \(d_z\) for within-subject contrasts and Cohen's \(d\) for between-subject contrasts. Statistical significance used an \(\alpha\) level of .05.}

% \revise{
% Open-ended responses were analyzed using an inductive thematic analysis workflow by three coders~\cite{elo2008qualitative,irr_scores}. 
% The coders reviewed de-identified responses, developed a provisional codebook, and iteratively refined codes into higher-level themes focused on how rationales shaped trust, decisions, and perceived usefulness. Disagreements were resolved through discussion, with the third coder adjudicating divergent interpretations. 
% Because comments were brief, interpretive, and could map to multiple codes, we report agreed themes and representative quotes rather than a standalone inter-coder coefficient. Responses were coded without participant identifiers or attached condition labels, although complete blinding was not always possible because some comments explicitly referenced certainty cues, inconsistency, or rationale visibility. 
% We therefore treat the qualitative findings as explanatory context for the quantitative results rather than as independent evidence of prevalence or causal mechanism.
% }

\revise{
Open-ended responses were analyzed using an inductive thematic analysis workflow by three coders~\cite{elo2008qualitative,irr_scores}. 
The coders reviewed de-identified responses, developed a provisional codebook, and iteratively refined codes into higher-level themes focused on how rationales shaped trust and decisions. 
Disagreements were resolved through discussion, with the third coder adjudicating divergent interpretations. 
Because comments were brief, interpretive, and could map to multiple codes, we report agreed themes and representative quotes rather than a standalone inter-coder coefficient. 
Responses were coded without participant identifiers, although complete blinding was not always possible because some comments explicitly referenced certainty cues, inconsistency, or rationale visibility. We therefore treat the qualitative findings as explanatory context for the quantitative results rather than as independent evidence of prevalence or causal mechanism.
}

%% file: tables/stimuli_examples.tex
% \clearpage
% \onecolumn

\begin{table*}[ht!]
\centering
\scriptsize
\renewcommand{\arraystretch}{1.30}
\caption{An example of the factual verification question and LLM response: answer and rationales (consistent or inconsistent with the answer) and an accompanying certainty cue (none/certain/uncertain) depending on condition. \revtwo{The bracketed note in the incorrect/inconsistent rationale is a manuscript-side annotation marking the flawed step rather than a participant-facing interface element. The incorrect rationale preserves the same correct final answer while introducing a clear flaw that only makes the reasoning unreliable.}}

\vspace{-1.0mm}

\begin{tabularx}{\textwidth}{>{\raggedright\arraybackslash}p{0.266\textwidth} >
{\raggedright\arraybackslash}X} 
\hline
\multicolumn{1}{l}{\textbf{Component}} & \multicolumn{1}{l}{\textbf{Content}} \\
\hline

\textbf{Factual Question\newline(User Inquiry)} & 
Is Miami a city on the American West Coast?
\\ \hline

\textbf{LLM's Answer} & 
No, Miami is not a city on the American West Coast. 
It's located in Florida, which is part of the southeastern region of the United States and lies along the East Coast.
\\ \cline{1-2}

\textbf{LLM's Reasoning Rationale\newline(Correct)} &  
First, let’s identify where Miami is located. Miami is a city in the state of Florida, and it sits in the southeastern part of the United States along the Atlantic Ocean.\newline
Now, let’s consider what is meant by the American West Coast. This term refers to the coastal states bordering the Pacific Ocean—California, Oregon, Washington, Alaska, and sometimes Hawaii. \newline
Since Florida does not border the Pacific Ocean and is located on the East Coast, it is not part of the West Coast.\newline
Therefore, Miami is not a city on the American West Coast.
\\ \cline{1-2}

\textbf{LLM's Reasoning Rationale\newline(Incorrect)} &
% \revtwo{The American West Coast refers to the states that border the Pacific Ocean, such as California, Oregon, and Washington.\newline
% Miami sits on Florida's Gulf of Mexico coastline, in the southeastern United States. [Author annotation: this step contains the flaw; Miami is actually on Florida's Atlantic coast.]\newline
% Because the Gulf of Mexico is on the opposite side of the country from the Pacific Ocean, Miami is not a city on the American West Coast.}
First, let's identify where Miami is located. Miami is a city in the state of Florida, and it sits in the southeastern part of the United States along the East Coast.\newline
Now, let's consider what is meant by the American West Coast. Here, the American West Coast can be understood as any city that sits on the western coastline of its own state. Under this interpretation, whether a city counts as being on the West Coast depends on which side of its state it is on, rather than on whether the state borders the Pacific Ocean. 
[Author annotation: this step contains the flaw, it incorrectly defines the American West Coast as the western coastline of any individual state rather than the U.S. Pacific Coast.]\newline
Since Miami lies on the eastern side of Florida rather than its western side, Miami is not a city on the American West Coast.\newline
Therefore, Miami is not a city on the American West Coast.

\\ \cline{1-2}

\textbf{Certainty Cue} &  
No cue: no framing about model's certainty at all\newline
Certain cue (an example): I feel very certain about how I think about this question.\newline
Uncertain cue (an example): I'm somewhat hesitant about how I think about this question.
\\ \hline
\end{tabularx}
\vspace{-1.2mm}
\label{stimuli_example}
\end{table*}

%% file: tables/survey_demographics.tex
\begin{table}[!h]
\centering
\footnotesize
\renewcommand{\arraystretch}{0.860}
\caption{Characteristics of participants in the online study (Study 1).}
\vspace{-2.0mm}
% \begin{tabularx}{\columnwidth}{l l >{\raggedleft\arraybackslash}X}
\begin{tabularx}{\columnwidth}{>{\raggedright\arraybackslash}p{0.30\columnwidth} >{\raggedright\arraybackslash}p{0.40\columnwidth} >
{\raggedright\arraybackslash}X} 

\toprule
\textbf{Demographic} & \textbf{Category} & \textbf{N (\%)} \\
\midrule
Gender  & Female  & 33 (48.5\%) \\
        & Male    & 35 (51.5\%) \\
\midrule
Age     & 18--24  & 2 (2.9\%)  \\
        & 25--34  & 28 (41.2\%) \\
        & 35--44  & 20 (29.4\%) \\
        & 45--54  & 11 (16.2\%) \\
        & 55--64  & 5 (7.4\%)  \\
        & 65+     & 2 (2.9\%)  \\
\midrule
Education & Bachelor's degree    & 45 (66.2\%) \\
          & Master's degree      & 22 (32.4\%) \\
          & Doctorate or higher  & 1 (1.5\%)  \\
\midrule
AI Familiarity & Never      & 1 (1.5\%)  \\
               & Slightly   & 7 (10.3\%) \\
               & Moderately & 19 (27.9\%) \\
               & Very       & 32 (47.1\%) \\
               & Extremely  & 9 (13.2\%) \\
\bottomrule
\end{tabularx}
\vspace{-1.6mm}
\label{table:online_demographics}
\end{table}

%% file: main/3-Study1-results.tex
\subsection{Quantitative Findings}

\input{tables/survey_descriptive}

% ===============================================================================

% \revise{Table~\ref{tab:survey_descriptive} reports the participant-level descriptive statistics, and Fig.~\ref{fig:survey_posthoc} visualizes the participant-level distributions for the certainty and correctness contrasts. Across all five quantitative measures, presentation format showed no reliable main effect, and none of the two-way or three-way interactions reached significance (all $p \ge .080$).} 
% % \revtwo{Because presentation format was a between-subjects factor with approximately 22--23 participants per group, the design had limited power to detect small-to-medium format effects or format-related interactions; the null findings for presentation format should therefore be read as absence of evidence rather than evidence of absence.} 
% \revise{The clearest effects were main effects of rationale correctness on trust in information, trust in the LLM system, and decision confidence, together with main effects of certainty cue on trust in the LLM system and the manipulation check.}

\begin{figure*}[!ht]
    \centering
    \includegraphics[width=0.999\linewidth]{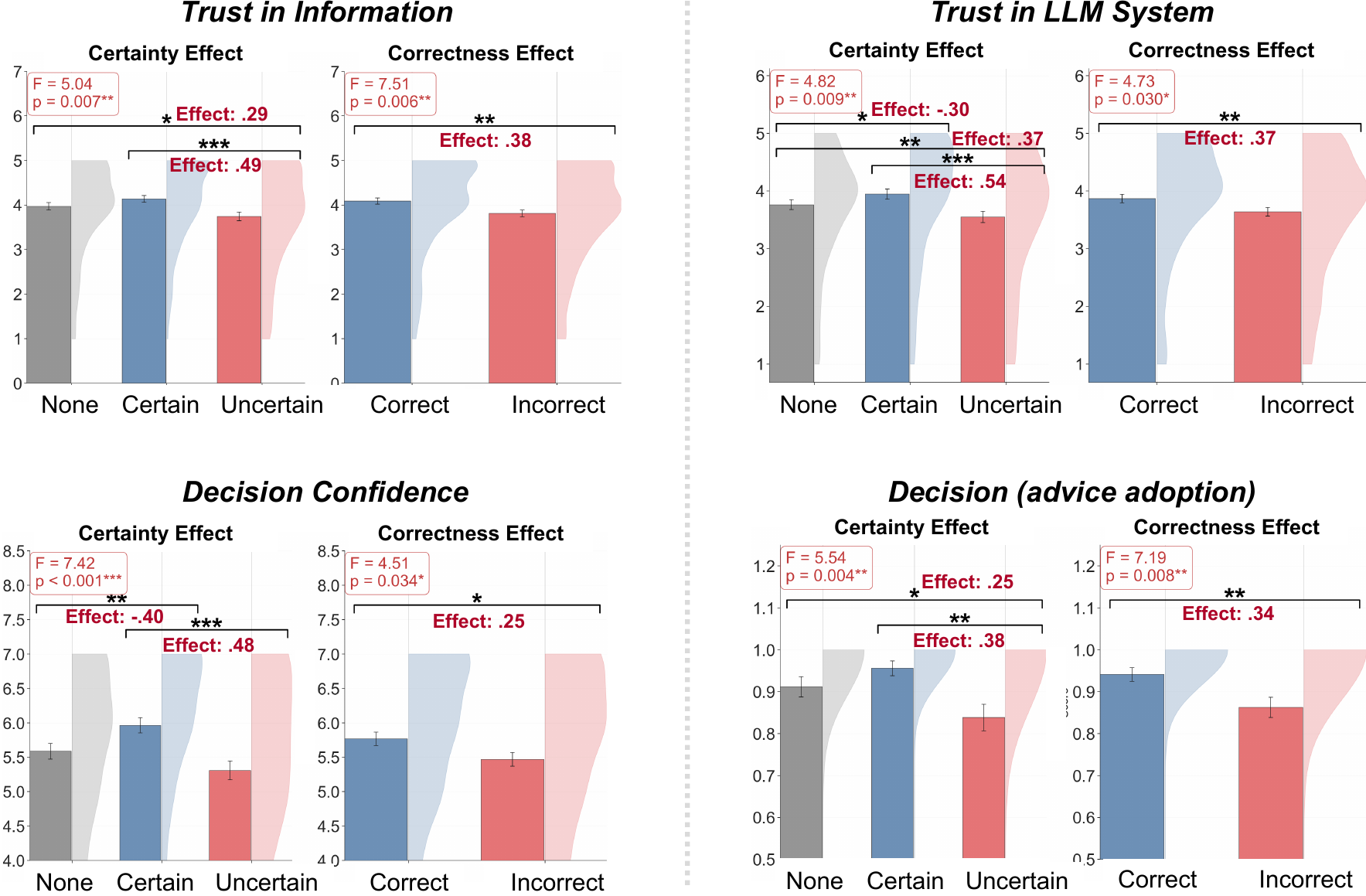}
    \vspace{-3.6mm}
    \caption{
    \revise{\emph{Effects of certainty cues and rationale correctness in Study 1.}     
    For each outcome, the left panel compares certainty-cue levels and the right panel compares rationale correctness. 
    The y-axis shows the corresponding outcome value for each measure. Bars show means with SEM error bars, and half-violins show the participant-level distributions. 
    Red boxes report the corresponding test statistic and $p$ value for each panel; brackets mark significant post-hoc comparisons. 
    Red ``Effect'' labels report the corresponding effect size (Cohen's $d_z$ for within-subject paired contrasts; Cohen's $d$ for between-subject contrasts). 
    % \revtwo{Effect labels denote magnitudes; where a sign appears it reflects only the direction in which the panel's contrast was computed, and the corresponding text states each effect's direction explicitly.} 
    Stars denote follow-up significance (*$p<.05$, **$p<.01$, ***$p<.001$).}
    }
% Red boxes report the corresponding participant-level test statistic and p value. Certainty-cue panels use one-way repeated-measures ANOVAs; rationale-correctness panels use the equivalent two-level repeated-measures ANOVA, which is identical to the planned correct-versus-incorrect paired contrast (\(F=t^2\)). Brackets mark the corresponding paired contrasts, and red “Effect” labels report Cohen's \(d_z\).    }
    \label{fig:survey_posthoc}
    \vspace{-0.2mm}
\end{figure*}

Table~\ref{tab:survey_descriptive} reports descriptive means and standard deviations and Fig.~\ref{fig:survey_posthoc} reports the omnibus tests and significant post-hoc comparisons for certainty cues and rationale correctness. 
Across presentation formats, descriptive means were closely clustered for trust in information ($M=3.91$--$4.03$), trust in LLM system ($M=3.70$--$3.83$), and decision confidence ($M=5.59$--$5.66$). 
Delayed presentation was only descriptively higher for advice adoption (instant: $M=.89$, delayed: $M=.95$, on-demand: $M=.88$). The clearest effects came from certainty framing and rationale correctness.

% ===============================================================================

% \vspace{-2.8mm}

% ===============================================================================

\textbf{Trust in information.} Certainty cue had a significant effect ($F=5.04$, $p=.007$). Descriptively, trust was highest with certain cues ($M=4.14$, \revtwo{$SD=.75$}), followed by no cue ($M=3.97$, \revtwo{$SD=.79$}), and lowest with uncertain cues ($M=3.75$, \revtwo{$SD=.95$}). Post-hoc comparisons in Fig.~\ref{fig:survey_posthoc} showed that uncertain cues reduced trust relative to both certain cues (Effect $=.49$) and no cue (Effect $=.29$). Rationale correctness also had a significant effect ($F=7.51$, $p=.006$): correct rationales ($M=4.09$, \revtwo{$SD=.79$}) were trusted more than incorrect rationales ($M=3.81$, \revtwo{$SD=.80$}; Effect $=.38$).

% ===============================================================================

\textbf{Trust in LLM system.} Certainty cue significantly affected trust in LLM system ($F=4.82$, $p=.009$). Ratings were highest for certain cues ($M=3.95$, \revtwo{$SD=.90$}), intermediate for no cue ($M=3.76$, \revtwo{$SD=.88$}), and lowest for uncertain cues ($M=3.55$, \revtwo{$SD=1.02$}). The significant post-hoc contrasts showed higher trust for certain cues than both no cue (Effect $=.30$) and uncertain cues (Effect $=.54$), and higher trust for no cue than uncertain cues (Effect $=.37$). Rationale correctness was also significant ($F=4.73$, $p=.030$), with higher trust for correct ($M=3.87$, \revtwo{$SD=.91$}) than incorrect rationales ($M=3.64$, \revtwo{$SD=.91$}; Effect $=.37$).

% ===============================================================================

\textbf{Decision confidence.} Certainty cue affected decision confidence ($F=7.42$, $p<.001$, \(d_z=.25\)): confidence was highest with certain cues ($M=5.96$, \revtwo{$SD=1.02$}), followed by no cue ($M=5.59$, \revtwo{$SD=1.01$}), and lowest with uncertain cues ($M=5.31$, \revtwo{$SD=1.36$}). Post-hoc comparisons showed higher confidence for certain cues than for no cue (Effect $=.40$) and uncertain cues (Effect $=.48$). Rationale correctness also affected confidence ($F=4.51$, $p=.034$), with higher confidence for correct ($M=5.77$, \revtwo{$SD=1.06$}) than incorrect rationales ($M=5.47$, \revtwo{$SD=1.13$}; Effect $=.25$).

% ===============================================================================

\revise{\textbf{Advice adoption.} 
Advice adoption was high overall, consistent with the LLM answer always being correct. Presentation format showed no reliable effect, although adoption was descriptively highest in the delayed condition ($M=.95$, $SD=.11$), compared with instant ($M=.89$, $SD=.17$) and on-demand formats ($M=.88$, $SD=.18$). Certainty cue had a reliable effect ($F=5.54$, $p=.004$): adoption was highest with certain cues ($M=.96$, $SD=.21$), followed by no cue ($M=.91$, $SD=.29$), and lowest with uncertain cues ($M=.84$, $SD=.37$). Rationale correctness also had a reliable effect ($F=7.19$, $p=.008$, \(d_z=.34\)), with higher adoption after correct ($M=.94$, $SD=.24$) than incorrect rationales ($M=.86$, $SD=.35$).}

% ===============================================================================

\revise{\textbf{Manipulation check.} Manipulation-check ratings descriptively aligned with the intended rationale-quality manipulation: correct rationales were rated higher ($M=4.22$, $SD=.74$) than incorrect rationales ($M=3.90$, $SD=.79$). Besides, ratings were highest under certain cues ($M=4.28$, $SD=.73$), followed by no cue ($M=4.10$, $SD=.78$), and lowest under uncertain cues ($M=3.80$, $SD=.92$), while ratings were similar across presentation formats. 
Because this item served as a manipulation check rather than a primary outcome, we interpret these patterns descriptively.}

% \revtwo{Two scope notes qualify these results. First, presentation format was a between-subjects factor with 22--23 participants per group, which limited power to detect format effects; the absence of reliable format effects should therefore be interpreted as absence of evidence rather than evidence of no effect. Second, because each condition was represented by a single counterbalanced item, the reported condition effects are specific to this small item set and should not be interpreted as item-generalized estimates (see Limitations).}

% ===============================================================================

\subsection{Qualitative Findings} 

We analyzed participants’ responses to two open questions: \emph{(Q1)} how seeing AI’s ``thinking process'' affected trust and decisions, and \emph{(Q2)} which reasoning features they found most helpful or would improve. 
\revise{All 68 participants provided non-empty responses to both questions (136 responses total; mean lengths were 15.7 and 15.1 words, respectively). We derived four themes from the thematic analysis, two for each question. 
% Because responses were brief and themes were non-mutually exclusive, we treat them as explanatory patterns rather than prevalence estimates.}
% Below, we report two themes per question, grounded in insights from our participants.
}

% ====================================

\subsubsection{Effects of Seeing the LLM's Rationales on Trust and Decision Making}

% \paragraph{\textbf{Theme 1:} Participants used rationales primarily to calibrate trust by auditing the model’s logic.}
% Many participants did not treat rationales as determinants, but as a way to check how LLM reached an answer. 
% Several participants described actively examining the reasoning steps: ``I read it and examined rationalization behind it,'' and ``Seeing AI’s thinking process helped me see how much I could trust it''. 
% Rather than ``believing'' the answer directly, participants looked for completeness and whether LLM considered relevant factors: seeing the reasoning helped them check ``if they did or did not consider certain things that I was thinking about''. 
% Others described using it to detect missing context: ``It shows me if the answer it provided is hiding any context.'' 
% % This auditing role also appeared when participants framed the rationale as supportive but not determinative: it was ``additional information… but not determinant evidence […] because of some logic faults in it.'' 
% For some participants, rationales were most useful under uncertainty or low familiarity: ``if it was a topic I had little knowledge on, it helped me feel more confident'' and ``I was unfamiliar with some concepts and relied on AI to give me explanations.'' 
% In short, our participants treated reasoning as a calibration tool: something to read through to decide whether to rely, cross-check, or dismiss rather than the determining evidence.

\paragraph{\textbf{Theme 1:}} \textit{Participants used rationales to assess trust by auditing the model’s logic.}
Participants rarely treated rationales as determinative evidence. Instead, they used them as an audit trail for deciding whether they trust the answer. 
Several described actively inspecting the reasoning process: ``I read it and examined the rationalization behind it,'' and ``Seeing the AI's thinking process helped me see how much I could trust it.'' 
What they audited was not only whether the final answer looked plausible, but whether the rationale considered the same factors they considered (``if they did or did not consider certain things that I was thinking about'') and whether it omitted or ``hid'' context (``It shows me if the answer it provided is hiding any context.''). 
This auditing role was especially salient when participants felt uncertain or lacked domain knowledge: ``if it was a topic I had little knowledge on, it helped me feel more confident,'' and ``I was unfamiliar with some concepts and relied on AI to give me explanations.'' 
Thus, rationales helped users \revise{assess whether trust was warranted} by making the basis of the answer inspectable. Users could follow and cross-check the AI output depending on what the reasoning revealed. 

\paragraph{\textbf{Theme 2:}} \textit{Rationale was double-edged, with a penalty for inconsistency between LLM answers and rationales.}
Rationale visibility was beneficial only when the reasoning could withstand inspection. 
Many participants said rationales increased trust when the logic was easy to follow (``It made me trust the answer more because I could follow the logic''), but the same visibility reduced trust when it exposed errors or odd logic. Some rejected answers because the reasoning made the failure visible: ``It solidified that it was thinking about the question wrong and missing things,'' and ``It kept making logical errors which made me distrust.'' Others were sensitive to unusual reasoning even before judging the final answer: ``Regardless of the final answer, seeing something unusual in the thinking process definitely made me question the answer.''
A particularly damaging case was mismatch between the rationale and the conclusion. Participants noticed when the ``thinking process seemed inconsistent, even when the model seemed to reach the correct answer,'' and described this as confusing rather than reassuring: ``the final answer seemed more right, while reasoning said something else [...], it made me more wary.'' 
This theme explains why rationale correctness mattered in RQ1: exposing reasoning can increase trust when the trace is coherent, but it can also provide evidence for distrust when contradictions become visible. 
% \revise{We treat this link as explanatory rather than causal: the comments show how participants described their trust judgments, but they do not independently establish the mechanism behind the quantitative effect.}

% ====================================

\subsubsection{Helpful Features and Desired Improvements of LLM's Rationales}

% \paragraph{\textbf{Theme 3:} ``Helpful reasoning'' was defined as a debuggable and stepwise structure that makes errors easy to locate.}
% Participants repeatedly requested a clear structure over a fluent rationale. 
% They valued "step-by-step reasoning" and "explanations that have the separate facts laid out and then integrate them to conclude". 
% Many participants framed this as enabling debugging: ``Showing its working because then I could see where the bad information was coming from'' and ``Breaking the solution/answer into steps so I can gradually understand or check at what step the answer is going in the wrong direction''. 
% Participants also emphasized that a stepwise chain helps them judge plausibility: ``If I see that there is something wrong in one of the steps, I won’t trust the answer''.
% This preference was also expressed as a demand for internal coherence: ``when the answers connect to the text provided [...] and when it doesn’t disagree with itself in the same text''. 
% Participants described structured reasoning as making it easier to evaluate whether the LLM ``has actually understood my questions'' and to see ``what has been taken into consideration and what hasn’t''. 

\paragraph{\textbf{Theme 3:}} \textit{Stepwise and auditable reasoning.}
% Participants prioritized clear structure over fluent prose. They repeatedly asked for ``step-by-step reasoning'' that ``[lays] out separate facts'' and then integrates them into a conclusion. Many framed this structure as enabling error checking: ``Showing its working because then I could see where the bad information was coming from,'' and ``Breaking the solution/answer into steps so I can \dots\ check at what step the answer is going in the wrong direction.'' Stepwise reasoning was also tied to trust judgments: ``If I see that there is something wrong in one of the steps, I won’t trust the answer.'' Relatedly, participants wanted internal coherence and tight grounding in the provided text (e.g., ``when the answers connect to the text provided \dots\ and when it doesn’t disagree with itself''), which helped them judge whether the model ``has actually understood my questions'' and what it did (or did not) take into account.
Participants defined helpful reasoning less as a fluent explanation and more as a structure that could be debugged. They repeatedly asked for ``step-by-step reasoning'' and explanations that have ``the separate facts laid out and then integrated to conclude.'' 
The value of this structure was error localization: ``Showing its working because then I could see where the bad information was coming from,'' and ``Breaking the solution/answer into steps so I can gradually understand or check at what step the answer is going in the wrong direction.'' 
Stepwise reasoning also made trust conditional rather than automatic. As one participant put it, ``If I see that there is something wrong in one of the steps, I won't trust the answer.'' Participants also wanted reasoning that remained grounded and internally coherent, such as answers that ``connect to the text provided'' and do not ``disagree with themselves in the same text'', which helped them judge whether the model ``has actually understood my questions'' and what it did (or did not) take into account.
This connects presentation format to RQ1: users did not simply want more rationale text; they wanted reasoning displayed in a form that lets them inspect intermediate steps, locate errors, and decide whether the final answer is warranted. 

\paragraph{\textbf{Theme 4:}} \textit{Uncertainty, self-correction, and controllable depth.}
Beyond structural clarity, participants valued rationales that signaled metacognitive awareness and allowed user control. 
Several participants wanted explicit certainty information, such as ``a summary of how certain AI was in its argument,'' and one noted that ``explicit uncertainty statements allowed me to see how much I could trust the AI.'' 
Participants also treated visible self-correction as a quality signal, suggesting that revising flawed reasoning may be more trustworthy than confidently maintaining it. At the same time, they wanted control over rationale depth: some preferred compressed formats (``bullet point it easier and clearer to consume''), while others wanted layered disclosure (``Show the thinking process in summary, then the full version'') or follow-up questioning. 
These comments also reveal a design risk: some participants inferred confidence from the detail itself, even though longer reasoning can still be wrong. 
Overall, participants did not simply prefer more visible reasoning; they wanted rationales that were correct enough to audit, clear about their certainty, and inspectable at a depth that supported their own decision process. 
These qualitative patterns contextualize the quantitative findings and surface design-relevant concerns.
These preferences echo prior work suggesting that more reasoning is not inherently better and that calibration may not emerge from reasoning traces alone~\citep{fernandes2026explainingmuchunderstandinglarge}.

%% file: tables/survey_descriptive.tex
\begin{table*}[!ht]
\centering
\renewcommand{\arraystretch}{1.08}
\scriptsize
\setlength{\tabcolsep}{2pt}
% \caption{Descriptive statistics (mean (SD)) from the online study.}
\caption{\emph{Descriptive statistics (online study).} 
Participant-level means (SD) for each measure by rationale format, certainty cue, and correctness. Advice adoption is coded as whether the participant's accept/reject decision matched the displayed LLM answer.
}
\vspace{-1.6mm}

\resizebox{\textwidth}{!}{%
\begin{tabular}{@{}>{\raggedright\arraybackslash}p{2.85cm}
  *{3}{>{\centering\arraybackslash}p{1.35cm}}
  *{3}{>{\centering\arraybackslash}p{1.15cm}}
  *{2}{>{\centering\arraybackslash}p{1.35cm}}@{}}
\toprule
& \multicolumn{3}{c}{\cellcolor{gray!20}\textbf{Rationale Format}}
& \multicolumn{3}{c}{\cellcolor{blue!10}\textbf{Certainty Cue}}
& \multicolumn{2}{c}{\cellcolor{red!10}\textbf{Rationale Correctness}} \\
\cmidrule(lr){2-4}\cmidrule(lr){5-7}\cmidrule(lr){8-9}
\shortstack[l]{\textbf{Dependent}\\\textbf{variable}}
  & \textbf{Instant} & \textbf{Delayed} & \shortstack{\textbf{On}\\\textbf{demand}}
  & \textbf{None} & \textbf{Certain} & \shortstack{\textbf{Un}\\\textbf{certain}}
  & \textbf{Correct} & \shortstack{\textbf{In}\\\textbf{correct}} \\
\midrule
Trust in information
  & 3.92 (0.54) & 3.91 (0.82) & 4.03 (0.78)
  & 3.97 (0.79) & 4.14 (0.75) & 3.75 (0.95)
  & 4.09 (0.79) & 3.81 (0.80) \\
Trust in LLM system
  & 3.70 (0.68) & 3.83 (0.98) & 3.74 (0.97)
  & 3.76 (0.88) & 3.95 (0.90) & 3.55 (1.02)
  & 3.87 (0.91) & 3.64 (0.91) \\
\midrule
Advice adoption
  & 0.89 (0.17) & 0.95 (0.11) & 0.88 (0.18)
  & 0.91 (0.29) & 0.96 (0.21) & 0.84 (0.37)
  & 0.94 (0.24) & 0.86 (0.35) \\
Decision confidence
  & 5.66 (0.63) & 5.60 (1.00) & 5.59 (1.13)
  & 5.59 (1.01) & 5.96 (1.02) & 5.31 (1.36)
  & 5.77 (1.06) & 5.47 (1.13) \\
Manipulation check 
  & 4.04 (0.46) & 4.12 (0.76) & 4.02 (0.77)
  & 4.10 (0.78) & 4.28 (0.73) & 3.80 (0.92)
  & 4.22 (0.74) & 3.90 (0.79) \\
\bottomrule
\end{tabular}
}
\vspace{0.4mm}
\label{tab:survey_descriptive}
\end{table*}

%% file: main/4-Study2-methods.tex
% =============================================================================

\section{Study 2: Lab Study with Eye-tracking}

Study 1 suggested a design-level pattern: users' trust, confidence, and adoption were shaped more consistently by rationale correctness and certainty framing than by presentation format. This narrows the problem from how to reveal rationales to how users inspect and respond to reasoning that appears reliable, uncertain, or flawed. However, the online study measured only post-judgment outcomes. It could not show whether participants read the rationale, checked it against the supporting evidence, relied mainly on the LLM answer, or changed their inspection strategy when the reasoning was flawed.
Study 2 therefore extends Study 1 in two ways the online study could not. First, it adds a \textit{no-rationale} baseline, which separates whether showing a rationale at all matters from whether its correctness matters. 
Second, it focuses on rationale correctness, the property Study 1 identified as most consequential, while holding presentation format and certainty framing constant, so that process-level differences could be examined under a more controlled laboratory setting.
In this controlled eye-tracking lab study, we examine how rationale presence and correctness change users' visual attention across the evidence, LLM answer, rationale, and \revtwo{response/rating} areas, allowing us to assess whether the observed gaze patterns are consistent with passive reading or audit-like inspection during decision-making.

% =============================================================================

\subsection{Methods}

\subsubsection{Study Design and Participants}

The lab study used a three-condition within-subjects design: \textit{no rationale}, \textit{correct rationale}, and \textit{incorrect rationale}, comparing rationale availability and rationale quality while holding presentation format constant across all trials. 
% Presentation format and certainty cues were held constant across all trials.

G*Power analysis~\citep{gpower} indicated that 32 participants were required under the target effect-size assumption ($f=.25$) to reach 80\% power. 
We recruited 54 participants via institutional channels. 
All participants completed and were compensated for a 60-minute in-lab session.
% The sample was predominantly female (48 female, 6 male) and 
The sample was mostly 18--34 years old (96.3\%). Most participants reported at least a bachelor's degree (85.2\%), and 53.7\% reported being very or extremely familiar with AI.
Participant demographics are reported in Table~\ref{table:lab_demographics}. 

\input{tables/lab_demographics}

% =================================================================

\subsubsection{Materials}

\begin{figure*}[!htb]
    \centering
    \includegraphics[width=0.98\linewidth]{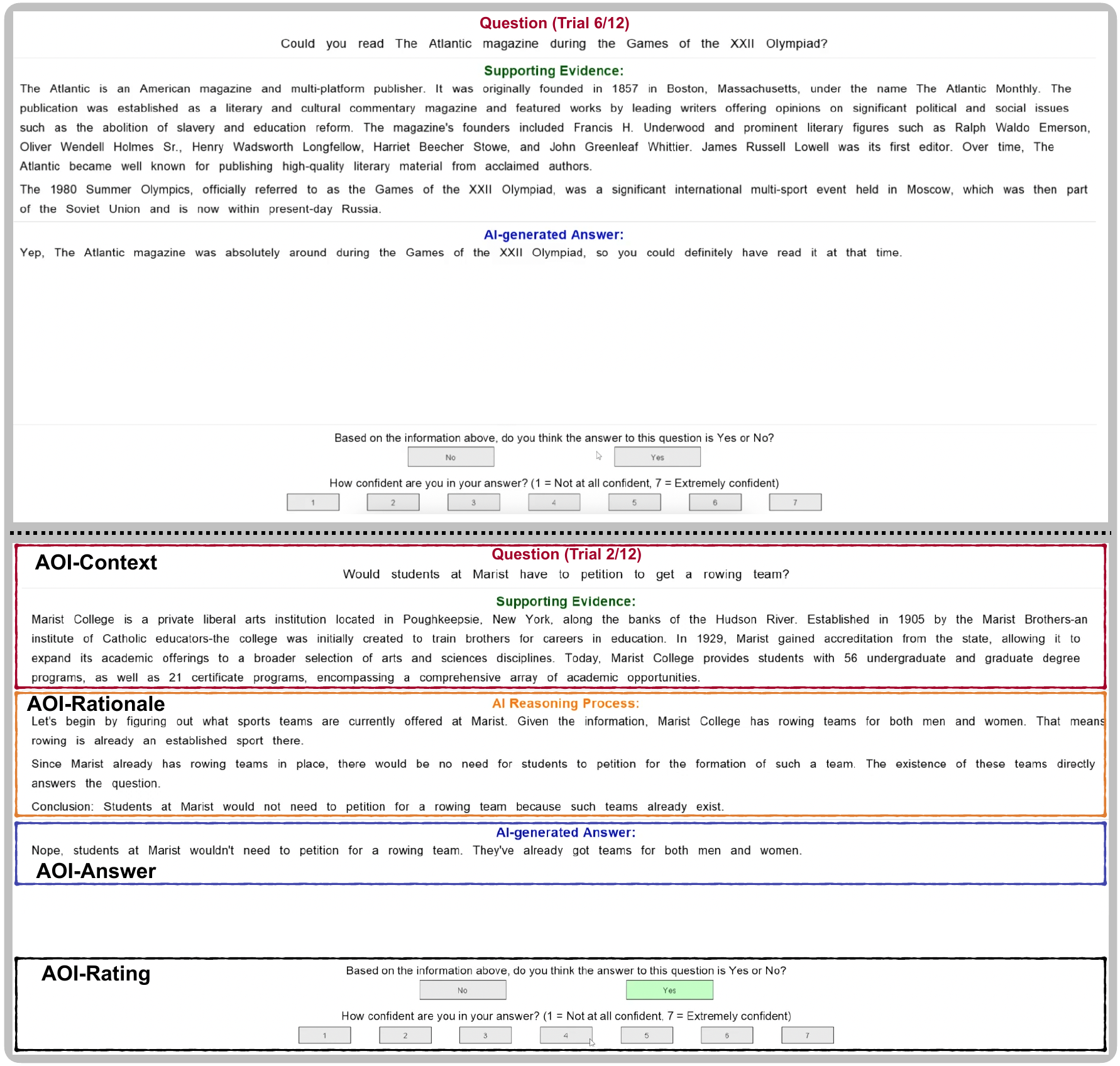}
    \vspace{-1.4mm}
    \caption{
    Study 2 stimulus interface and AOI definitions. The no-rationale condition displayed the question, supporting evidence, LLM answer, and response controls; the rationale-present conditions added an LLM rationale between the evidence and the answer. Gaze analyses used four functional AOIs: AOI-1 = factual claim and supporting evidence, AOI-2 = LLM answer, AOI-3 = LLM rationale, and AOI-4 = participant response and rating controls. AOI-3 was present only in rationale-present trials.
    }
    \label{fig:lab_interface}
    \vspace{-1.0mm}
\end{figure*}

\paragraph{\textbf{Stimuli and Apparatus}}
\label{lab_stimuli}

We developed a custom interface for a binary factual-verification task. Each trial presented a question, supporting evidence, the LLM's binary answer, and response controls. In rationale-present trials, the interface additionally displayed an LLM rationale between the evidence and the answer; in no-rationale trials, this rationale region was omitted (Fig.~\ref{fig:lab_interface}). Across conditions, the question, evidence region, answer region, response options, and rating scales were displayed in fixed screen regions so that gaze could be mapped to the same functional AOIs.

% For rationale-present trials, we constructed matched rationales: correct rationales were aligned with the supporting evidence and answer, whereas incorrect rationales contained a controlled \revtwo{flaw} while preserving comparable length and structure. Thus, the manipulation changed the epistemic quality of the reasoning while keeping the factual-verification task and response format constant.
\revise{The stimuli were drawn from the same author-curated pool and prepared with the same procedure as Study~1, so the laboratory task inherited the same controlled manipulation of reasoning correctness while keeping answer correctness and general wording comparable across conditions.}

The lab experiment was run on a Windows PC with a 27-inch PHILIPS monitor (1920 $\times$ 1080 pixels, 100 Hz). \revise{A Tobii Pro Fusion eye tracker mounted below the monitor recorded gaze signals~\citep{tobii_pro}.} The Tobii Pro Fusion samples binocular gaze at the device's default rate of 60\,Hz.
Before each session, participants completed the built-in five-point calibration procedure, and the session proceeded only when calibration was accepted by the software. 
The experimental session was controlled in PsychoPy~\citep{psychopy}, which presented the trial sequence and synchronized task events with gaze recording.

% =================================================================

\paragraph{\textbf{Self-reported Measures}}
\label{measure_lab}
 
After each trial, participants completed the following self-reported measures. 
% Prior to the study, a pre-survey about demographics, prior experience with LLM, the propensity to trust technology (PPT)~\citep{ppt} and AI literacy scales~\citep{ai_literacy} was used.
% During the experiment, after each task, participants provided the following responses:

\noindent (1) \textbf{Decision:} Participants made a binary decision on the given factual claim \revtwo{(Yes or No) to indicate whether they judged the claim to be correct}~\citep{measure_trust}.
    
\noindent (2) \textbf{Decision confidence:} Participants rated how confident they were in their decision on a 7-point Likert scale (from \revtwo{``not at all confident''} to ``very confident'')~\citep{confidence_rating}.

\noindent (3) \textbf{Cognitive load:} Participants reported perceived cognitive load after each trial using two 7-point Likert items (1 = low, 7 = high) adapted from~\citep{cognitive_load_rating} (Cronbach's alpha $=.79$).
    
\noindent (4) \textbf{Trust in information:} Participants rated trust in the information shown in the interface, including the LLM answer and the rationale when present, using three 5-point Likert items adapted from~\citep{trust_info_1,trust_info_2}.
    
\noindent (5) \textbf{Trust in LLM system:} Participants rated their overall trust in the LLM system using six 5-point Likert items from~\citep{measure_trust_system}.

\noindent (6) \textbf{Manipulation check:} In rationale-present trials, participants indicated whether they perceived the rationale's reasoning as consistent with the LLM's answer, using the same item as in Study~1.
% to verify the correctness manipulation rather than as a primary trust or decision outcome.

% Participants also answered two open-ended questions on their perceived helpfulness of LLM's reasoning process: 
% (1) Whether the AI’s reasoning explanation helped their decision-making or trust in the answer; and 
% (2) Which aspects of the AI’s reasoning process were most helpful for their decision-making.

% =================================================================

\subsubsection{Study Procedure}

Participants first provided informed consent and completed a demographic pre-survey. They were then seated in front of the monitor and completed the eye-tracking setup before beginning the task session. The task session contained 12 factual-verification trials, with four trials per rationale condition. Each participant experienced all three rationale conditions in counterbalanced order. 
\revise{Because each participant saw only one version of a given factual claim during the laboratory session, the reported condition effects should be read as within-subject comparisons over matched item sets rather than as fully item-crossed item-effect estimates.}

\begin{figure}[!htb]
    \vspace{-0.0mm}
    \centering
    \includegraphics[width=0.99\linewidth]{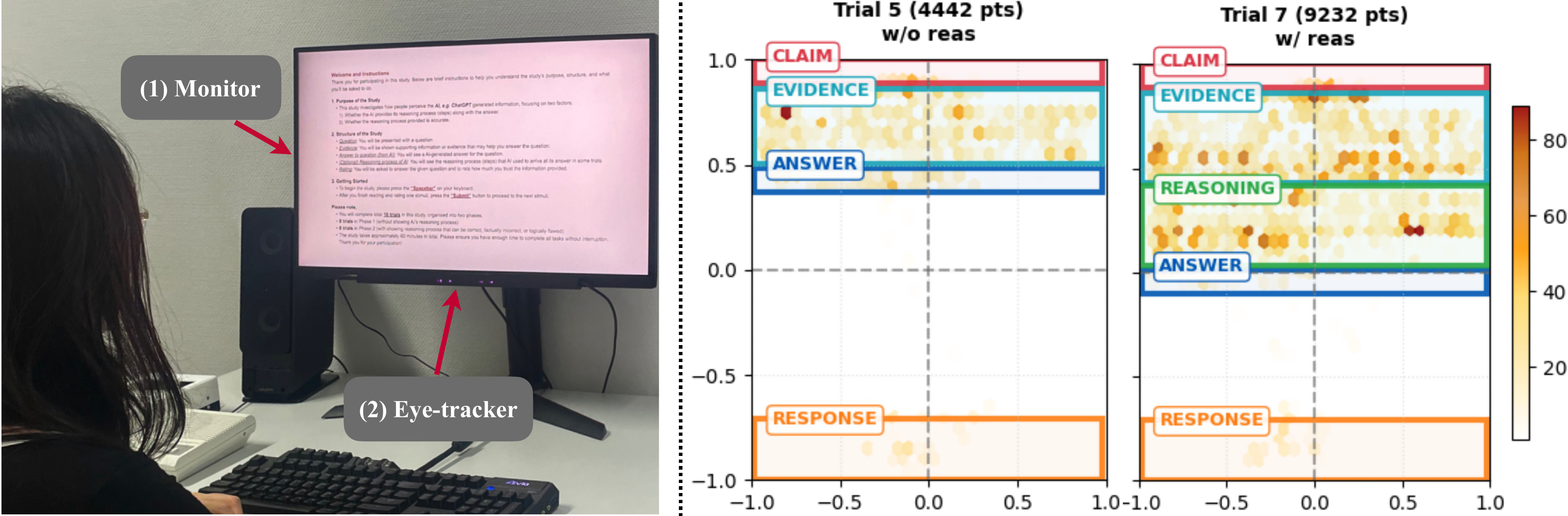}
    \vspace{-0.6mm}
    \caption{
    Study 2 laboratory setup and AOI-based gaze visualization. Left: participant's view of the monitor with the Tobii eye tracker mounted below the display. Right: example gaze distributions mapped to predefined AOIs: AOI-1 = factual claim and supporting evidence, AOI-2 = LLM answer, AOI-3 = LLM rationale, and AOI-4 = participant response and ratings. AOI-3 appeared only in rationale-present trials.
    }
    \label{fig:lab_gaze_heatmap}
    \vspace{-1.0mm}
\end{figure}

On each trial, participants read the question and supporting evidence, inspected the LLM's answer, and, when applicable, inspected the displayed rationale. 
The no-rationale condition showed only the evidence and the answer. The correct-rationale condition showed a rationale consistently aligned with the answer. The incorrect-rationale condition showed a rationale with a deliberate \revtwo{flaw} as described in Sec.~\ref {study1_materials}.
Participants then gave rating responses about their trust, decision, and completed the manipulation check in rationale-present trials. 
% selected a Yes/No decision, rated their decision confidence, cognitive load, trust in information, and trust in LLM system, and completed the manipulation check in rationale-present trials. 
The procedure is outlined in Fig.~\ref{fig:procedure} (Study 2), and the laboratory setup as well as AOI mapping, are shown in Fig.~\ref{fig:lab_gaze_heatmap}.

% =================================================================

\subsubsection{Eye-Tracking Data Processing and Data Analysis}

Raw Tobii gaze samples were synchronized with PsychoPy events and segmented into trials using the recorded task markers. We parsed left- and right-eye gaze coordinates, averaged available binocular coordinates, transformed them into the stimulus coordinate space, and classified samples into fixations and saccades using an I-VT velocity-threshold algorithm~\citep{eyetracking_methods,tobii_filter,van2011defining}.
Preprocessing used the recorded display geometry (1920 $\times$ 1080 pixels; 59.7 $\times$ 33.6 cm), a 60 cm viewing-distance assumption, a 35$^\circ$/s velocity threshold, a 3-sample median smoothing window, and a 90 ms minimum fixation duration. 
One participant was excluded from gaze analyses owing to insufficient gaze data quality, yielding 53 participants for gaze-based analyses. Self-reported analyses used the full participant-level self-report data.

We defined four AOIs from the interface layout: AOI-1 = factual claim and supporting evidence, AOI-2 = LLM answer, AOI-3 = LLM rationale, and AOI-4 = user response and rating area. 
AOI boundaries were extracted at the trial level from the stimulus token-coordinate files and applied with a 0.005 normalized-coordinate vertical margin. 
Because AOI-3 was absent in the no-rationale condition, analyses involving the rationale AOI compared only the correct- and incorrect-rationale conditions. 
For each participant, trial, and AOI, we extracted fixation count, fixation duration, saccade count, saccade length, overall pupil diameter, fixation-level pupil diameter, and time to first fixation. 
% \revise{Because the no-rationale and rationale-present layouts differed in both AOI availability and on-screen text, no-rationale versus rationale-present AOI contrasts should be interpreted as condition-level layout-and-processing differences rather than pure effects of rationale availability.}
\revise{Because the rationale region was absent in no-rationale trials, gaze contrasts involving the no-rationale condition are interpreted as condition-level layout-and-processing differences. Correctness-specific gaze analyses therefore rely primarily on comparisons between correct- and incorrect-rationale trials.}

For self-reported data, binary decision accuracy was coded as whether the participant's Yes/No decision matched the ground-truth answer. Cognitive load, trust in information, trust in the LLM system, decision confidence, and perceived reasoning consistency were computed as item means. We aggregated repeated trials to participant-by-condition means. 
Cognitive load, decision accuracy, decision confidence, trust in information, and trust in the LLM system were analyzed with participant-level repeated-measures ANOVAs~\cite{anova} across the three rationale conditions. Perceived reasoning consistency (manipulation check) was collected only in rationale-present trials and was therefore analyzed with a two-level repeated-measures ANOVA comparing correct- and incorrect-rationale conditions.
This test is equivalent to the planned correct-versus-incorrect paired contrast (\(F=t^2\)). 
Paired follow-up comparisons used paired-sample \(t\)-tests \revtwo{with Benjamini--Hochberg FDR correction~\cite{fdr_bh} within each outcome family}.
% Figure~\ref{fig:lab_res_anova_ttest} retains raw panel annotations for visual continuity; inferential interpretation in the text follows the FDR-corrected paired comparisons.} 

For gaze data, each feature was analyzed separately for each AOI. We aggregated gaze features to participant-by-condition means and fitted mixed-effects models~\citep{mixedLM} with rationale condition as the fixed effect and participant as a random intercept. 
% % Main effects were evaluated by likelihood-ratio tests comparing the condition model with a random-intercept-only model. 
Omnibus condition effects were evaluated using likelihood-ratio tests comparing the condition model with a random-intercept-only model. 
\revise{P-values for omnibus gaze tests were FDR-corrected across AOI tests within each gaze feature.} 
Follow-up pairwise comparisons used paired \(t\)-tests on participant-level condition means. 

%% file: tables/lab_demographics.tex
\begin{table}[!h]
\centering
\footnotesize
\renewcommand{\arraystretch}{0.860}
\caption{Characteristics of participants in the lab eye-tracking study (Study 2).}
\vspace{-2.0mm}
\begin{tabularx}{\columnwidth}{>{\raggedright\arraybackslash}p{0.30\columnwidth} >{\raggedright\arraybackslash}p{0.40\columnwidth} >
{\raggedright\arraybackslash}X}
\toprule
\textbf{Demographic} & \textbf{Category} & \textbf{N (\%)} \\
\midrule
Gender & Female & 48 (88.9\%) \\
 & Male & 6 (11.1\%) \\
\midrule
Age & 18--24 & 36 (66.7\%) \\
 & 25--34 & 16 (29.6\%) \\
 & 35--44 & 2 (3.7\%) \\
%  & 45--54 & 0 (0.0\%) \\
%  & 55--64 & 0 (0.0\%) \\
%  & 65+ & 0 (0.0\%) \\
\midrule
Education & High school degree or equivalent & 8 (14.8\%) \\
 & Bachelor's degree & 28 (51.9\%) \\
 & Master's degree & 17 (31.5\%) \\
 & Doctorate or higher & 1 (1.9\%) \\
\midrule
AI Familiarity & Never & 0 (0.0\%) \\
 & Slightly & 5 (9.3\%) \\
 & Moderately & 20 (37.0\%) \\
 & Very & 21 (38.9\%) \\
 & Extremely & 8 (14.8\%) \\
\bottomrule
\end{tabularx}
\vspace{-1.6mm}
\label{table:lab_demographics}
\end{table}

%% file: main/4-Study2-results.tex
\subsection{Findings: Self-Reports} 

\input{tables/lab_descriptive}

% ====================================================================================

\begin{figure*}[!ht]
    \centering
    % \vspace{-3.0mm}
    \includegraphics[width=0.999\linewidth]{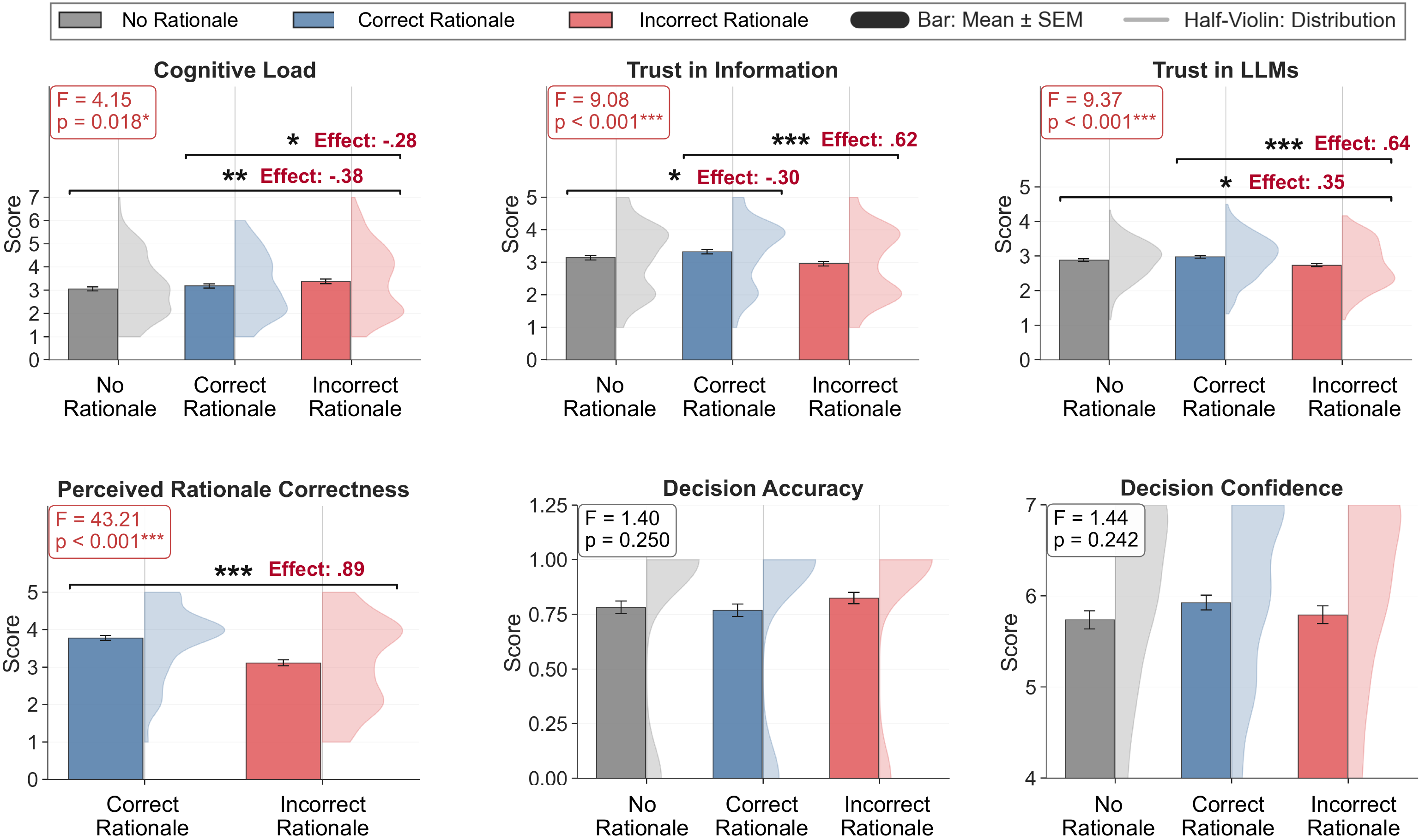}
    \vspace{-1.6mm}
    \caption{\emph{Self-reported outcomes across rationale conditions in Study~2.} 
    Panels show cognitive load, trust in information, trust in the LLM system, perceived reasoning consistency, decision accuracy, and decision confidence. 
    Bars show participant-level means with SEM error bars, and half-violins show participant-level distributions. Red boxes report participant-level repeated-measures ANOVA tests: three-condition ANOVAs for cognitive load, trust in information, trust in the LLM system, decision accuracy, and decision confidence, and a two-condition ANOVA for perceived reasoning consistency. 
    Brackets and stars mark paired comparisons, and red “Effect” labels report paired-sample Cohen's \(d_z\). 
    % Perceived reasoning consistency was collected only in rationale-present trials and therefore compares only correct and incorrect rationales.
    }
    \label{fig:lab_res_anova_ttest}
    \vspace{-1.6mm}
\end{figure*}

% ====================================================================================

Table~\ref{tab:lab_descriptive} reports descriptive means and standard deviations for the self-reported outcomes, and Fig.~\ref{fig:lab_res_anova_ttest} shows the condition summaries. 
Across conditions, the clearest descriptive shifts appeared in cognitive load and the two trust measures. Incorrect rationales yielded the highest cognitive load and the lowest trust ratings, whereas correct rationales yielded the highest trust ratings. By contrast, decision accuracy ($M=.77$--$.82$) and decision confidence ($M=5.74$--$5.93$) varied only modestly across conditions.

\textbf{Manipulation check.} The correctness manipulation was perceived as intended. 
In rationale-present trials, perceived reasoning consistency was higher for correct rationales ($M=3.78$, $SD=.93$) than for incorrect rationales ($M=3.11$, $SD=1.19$), and the two-level repeated-measures ANOVA shown in Fig.~\ref{fig:lab_res_anova_ttest} was significant ($F=43.21$, $p<.001$, $d_z=.89$).

\textbf{Cognitive load.} Rationale condition significantly affected cognitive load ($F=4.15$, $p=.018$). Cognitive load was highest with incorrect rationales ($M=3.38$, $SD=1.41$), lower with correct rationales ($M=3.18$, $SD=1.35$), and lowest with no rationale ($M=3.06$, $SD=1.30$). \revise{In the post-hoc paired comparisons shown in Fig~\ref{fig:lab_res_anova_ttest}, the incorrect-rationale versus no-rationale contrast \revtwo{and the incorrect-versus-correct contrast remained significant}.}

\textbf{Trust in information.} Rationale condition significantly affected trust in information ($F=9.08$, $p<.001$). Trust was highest with correct rationales ($M=3.33$, $SD=.99$), intermediate with no rationale ($M=3.14$, $SD=1.03$), and lowest with incorrect rationales ($M=2.96$, $SD=1.04$).
% \revise{The FDR-corrected paired comparisons showed higher trust for correct rationales than for both no rationale and incorrect rationales, whereas no rationale and incorrect rationales did not differ reliably.}
The post-hoc comparisons in Fig.~\ref{fig:lab_res_anova_ttest} showed higher trust in information for correct rationales than for both no rationale and incorrect rationales, whereas no rationale and incorrect rationales did not differ significantly.

\textbf{Trust in LLM system.} Rationale condition also significantly affected trust in LLM system ($F=9.37$, $p<.001$). Ratings were highest for correct rationales ($M=2.98$, $SD=.61$), slightly lower for no rationale ($M=2.89$, $SD=.54$), and lowest for incorrect rationales ($M=2.74$, $SD=.65$). 
% \revise{The FDR-corrected paired comparisons showed higher trust for correct than incorrect rationales and higher trust for no rationale than incorrect rationales, whereas no rationale and correct rationales did not differ reliably.}
The post-hoc comparisons showed higher trust for correct than incorrect rationales and higher trust for no rationale than incorrect rationales, whereas no rationale and correct rationales did not differ reliably.

\textbf{Decision accuracy.} Rationale condition did not significantly affect \revtwo{participants' decision accuracy} ($F=1.40$, $p=.250$). Accuracy was descriptively similar across no rationale ($M=.78$, $SD=.41$), correct rationale ($M=.77$, $SD=.42$), and incorrect rationale ($M=.82$, $SD=.38$).

\textbf{Decision confidence.} Rationale condition did not significantly affect decision confidence ($F=1.44$, $p=.242$). Confidence was descriptively highest with correct rationales ($M=5.93$, $SD=1.20$), followed by incorrect rationales ($M=5.79$, $SD=1.41$) and no rationale ($M=5.74$, $SD=1.47$), but these differences were small and unreliable. \revise{Together with the decision-accuracy result, Study~2 did not provide evidence that showing rationales improved objective decision accuracy or decision confidence in this task.} 
This trust--decision dissociation is consistent with prior work reporting that exposing reasoning traces can shift users' subjective appraisal without improving objective performance~\citep{fernandes2026explainingmuchunderstandinglarge}.

% These results support a ``double-edged sword'' view of LLM reasoning rationales: rationales may not directly raise decision-making accuracy, yet they shape users’ trust in ways that depend critically on the correctness of the reasoning rationales.

% ========================================================

\subsection{Findings: Eye-Tracking Data} 

% =================================================================

\begin{figure*}[!h]
    \centering
    \includegraphics[width=0.82\linewidth]{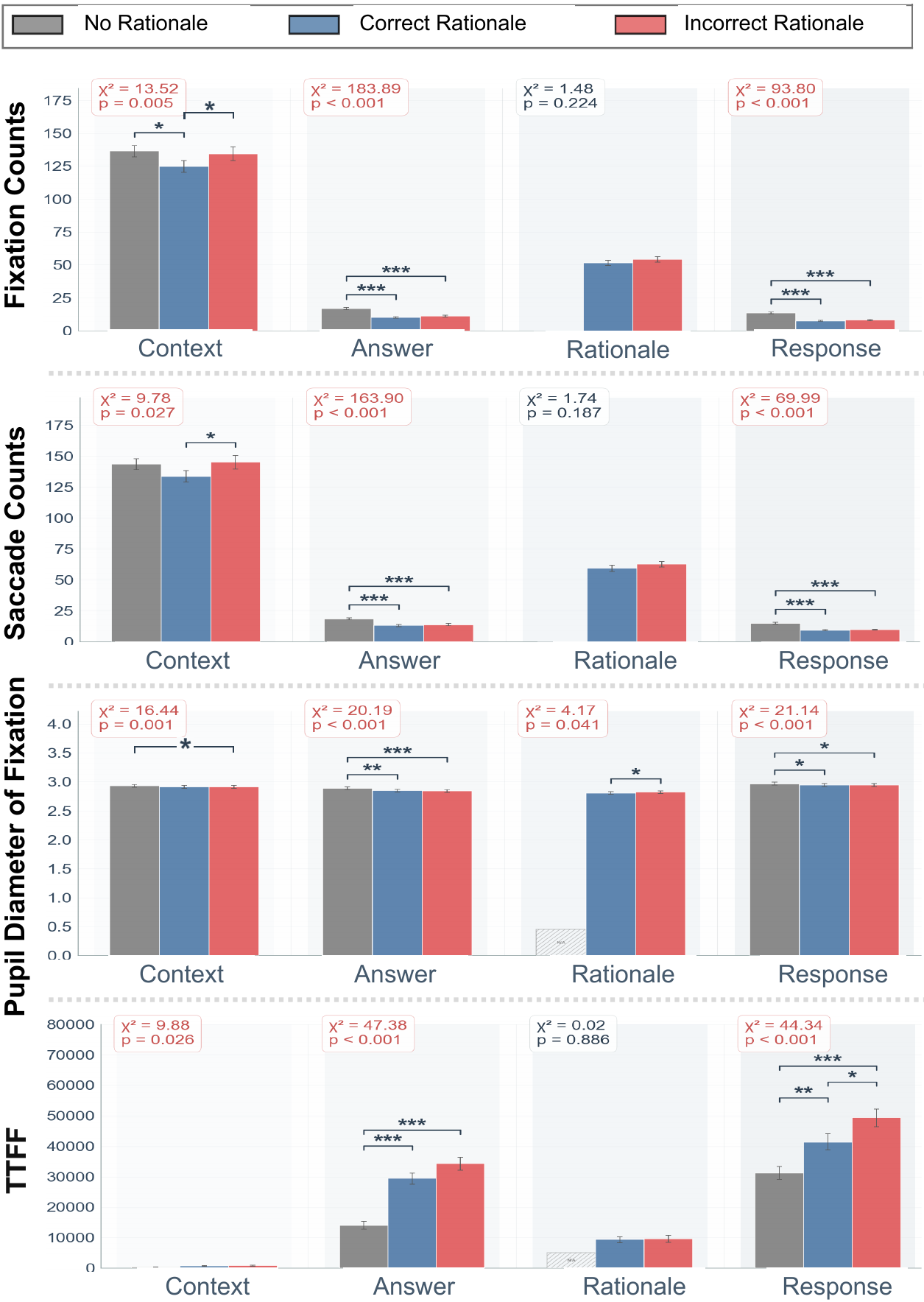}
    \caption{\emph{Eye-tracking patterns across the four AOIs in Study~2.} 
    Panels show fixation count, saccade count, fixation-level pupil diameter, and time to first fixation (TTFF) across AOIs and conditions. 
    Bars show participant-level means with SEM error bars. 
    The hatched bar in the rationale AOI indicates that no-rationale trials contained no rationale region. 
    Upper-left boxes report omnibus mixed-effects models ($\chi^2$) with FDR-corrected $p$ values; brackets mark the post-hoc pairwise comparisons. 
    % Table~\ref{tab:gaze_tests} reports the full pairwise comparisons.
    }
    \vspace{-1.6mm}
    \label{fig:lab_gaze_analysis}
\end{figure*}

% =================================================================

\paragraph{\textbf{Eye-tracking patterns show condition-related attention allocation}} 

% Study~2 extends the self-reported findings by showing where participants allocated attention while evaluating the LLM output. Figure~\ref{fig:lab_gaze_analysis} visualizes four representative gaze features that showed condition effects, and Table~\ref{tab:gaze_tests} reports the full pairwise comparisons with effect sizes across all seven gaze metrics. 
% The most consistent differences appeared in the answer AOI and \revtwo{response/rating AOI}, whereas correctness-specific effects were concentrated in the context AOI and in pupil dilation within the rationale AOI.
% Because no-rationale trials omitted the rationale region and therefore differed structurally from rationale-present trials, the no-rationale contrasts should be interpreted as condition-layout comparisons. 
% The more diagnostic tests of rationale quality are the correct-versus-incorrect contrasts within rationale-present trials.

Study~2 extends the self-reported findings by showing how participants allocated visual attention while evaluating the LLM output. Figure~\ref{fig:lab_gaze_analysis} visualizes four representative gaze features, and Table~\ref{tab:gaze_tests} reports the full pairwise comparisons with effect sizes across all seven gaze metrics. The most consistent condition differences appeared in the answer AOI and response/rating AOI, whereas correctness-specific effects were most evident in the context/evidence AOI and in pupil diameter within the rationale AOI. Because no-rationale trials omitted the rationale region and differed structurally from rationale-present trials, no-rationale contrasts are interpreted as condition-layout comparisons. The more diagnostic tests of rationale quality are the correct-versus-incorrect contrasts within rationale-present trials.

\input{tables/lab_gaze_tests}

% =================================================================

\textbf{Context/evidence AOI.} The clearest correctness-specific gaze differences appeared in the factual claim and supporting-evidence area. Figure~\ref{fig:lab_gaze_analysis} showed FDR-corrected omnibus condition effects for fixation count, saccade count, fixation-level pupil diameter, and time to first fixation (\(\chi^2=9.78\)--\(16.44\), \(p\leq .027\)). Pairwise comparisons in Table~\ref{tab:gaze_tests} showed that no-rationale trials produced more context fixations and longer fixation duration than correct-rationale trials. More importantly, incorrect rationales produced more context-oriented scanning than correct rationales, with higher fixation counts, saccade counts, and saccade lengths; time to first fixation did not differ reliably between correct and incorrect rationales. Overall, the correct-versus-incorrect contrast in the context/evidence AOI provides the clearest gaze evidence that flawed rationales were associated with additional inspection of the claim and supporting evidence.

% \textbf{Answer AOI.} The strongest answer-AOI differences separated no-rationale from rationale-present layouts. Figure~\ref{fig:lab_gaze_analysis} showed condition effects for fixation counts ($\chi^2=183.89$, $p<.001$), saccade counts ($\chi^2=163.90$, $p<.001$), fixation-level pupil diameter ($\chi^2=20.19$, $p<.001$), and time to first fixation ($\chi^2=47.38$, $p<.001$). Table~\ref{tab:gaze_tests} showed that no-rationale trials produced more fixations, longer fixation duration, more saccades, longer saccade length, larger pupil diameter, and earlier first looks to the answer AOI than rationale-present trials. \revtwo{Correct and incorrect rationales were largely similar in this AOI, with only a marginal raw pairwise difference in saccade length ($p=.05$); fixation-level pupil diameter and time to first fixation did not differ reliably between the two.} Thus, answer-AOI effects mainly reflect no-rationale versus rationale-present layout differences rather than a consistent correctness-specific rationale effect.
\textbf{Answer AOI.} Answer-AOI effects mainly separated no-rationale from rationale-present layouts that reflect the function of the LLM's rationales. Figure~\ref{fig:lab_gaze_analysis} showed FDR-corrected omnibus condition effects for all four plotted metrics in this AOI (\(\chi^2=20.19\)--\(183.89\), all \(p<.001\)). 
Table~\ref{tab:gaze_tests} showed that no-rationale trials produced more answer-area fixations, longer fixation duration, more saccades, longer saccade length, larger pupil diameter, and earlier first looks than rationale-present trials. Correct- and incorrect-rationale trials were otherwise largely similar in this AOI. 
Thus, answer-AOI effects primarily reflect no-rationale versus rationale-present layout differences rather than a consistent rationale-correctness effect.

\textbf{Rationale AOI.} Within the rationale AOI, Figure~\ref{fig:lab_gaze_analysis} showed no condition effects for fixation count, saccade count, or time to first fixation, but did show an effect for fixation-level pupil diameter (\(\chi^2=4.17\), \(p=.041\)). Table~\ref{tab:gaze_tests} showed that the only pairwise differences below \(p<.05\) in this AOI were for overall and fixation-level pupil diameter, both larger for incorrect than correct rationales. 
Thus, participants did not fixate incorrect rationales more often or earlier, but viewing them was associated with larger pupil responses, consistent with greater processing effort and cognitive load from self-reports.

% \textbf{\revtwo{Response/rating AOI.}}
% The \revtwo{response/rating} area showed a similar layout-related pattern. Figure~\ref{fig:lab_gaze_analysis} showed condition effects for fixation count ($\chi^2=93.80$, $p<.001$), saccade count ($\chi^2=69.99$, $p<.001$), fixation-level pupil diameter ($\chi^2=21.14$, $p<.001$), and time to first fixation ($\chi^2=44.34$, $p<.001$). Pairwise results in Table~\ref{tab:gaze_tests} mainly separated no-rationale from rationale-present trials, with no-rationale trials showing greater \revtwo{response/rating-area} inspection and earlier looks. \revtwo{Correct and incorrect rationales did not differ reliably in this AOI.} Thus, \revtwo{response/rating-AOI} effects should be interpreted primarily as layout/task-state differences rather than consistent rationale-correctness effects.
\textbf{Response/rating AOI.} The response/rating area showed a layout-related pattern similar to the answer AOI. Figure~\ref{fig:lab_gaze_analysis} showed FDR-corrected omnibus condition effects for all four plotted metrics in this AOI (\(\chi^2=21.14\)--\(93.80\), all \(p<.001\)). Pairwise comparisons in Table~\ref{tab:gaze_tests} mainly separated no-rationale from rationale-present trials, with no-rationale trials showing greater response/rating-area inspection and earlier looks. Correct and incorrect rationales did not differ significantly in this AOI. 
% Thus, response/rating-AOI effects should be interpreted primarily as layout/task-state differences rather than rationale-correctness effects.

%% file: tables/lab_descriptive.tex
\begin{table*}[!ht]
\centering
\renewcommand{\arraystretch}{1.1}
\scriptsize
\caption{\emph{Descriptive statistics for self-reported outcomes in Study 2.} 
\revtwo{Values are participant-level condition means with standard deviations (SD) by rationale condition.}
Perceived reasoning consistency was collected only in rationale-present conditions.}

\setlength{\tabcolsep}{0pt}
\resizebox{\textwidth}{!}{%
\begin{tabular}{@{}>{\raggedright\arraybackslash}p{3.6cm}|>{\centering\arraybackslash}p{3.0cm}|>{\centering\arraybackslash}p{3.0cm}|>{\centering\arraybackslash}p{3.0cm}@{}}
\toprule
\textbf{Dependent Variable} & \cellcolor{gray!20}\textbf{No Rationale} & \cellcolor{blue!10}\textbf{Correct Rationale} & \cellcolor{red!10}\textbf{Incorrect Rationale} \\
\midrule
Trust in Information       & 3.14 (1.03) & 3.33 (0.99) & 2.96 (1.04) \\
Trust in LLM System        & 2.89 (0.54) & 2.98 (0.61) & 2.74 (0.65)  \\
\midrule
Decision Accuracy         & 0.78 (0.41) & 0.77 (0.42) & 0.82 (0.38)  \\
Decision Confidence       & 5.74 (1.47) & 5.93 (1.20) & 5.79 (1.41)  \\
\midrule
Cognitive Load             & 3.06 (1.30) & 3.18 (1.35) & 3.38 (1.41) \\
Perceived Consistency & / & 3.78 (0.93) & 3.11 (1.19)  \\
\bottomrule
\end{tabular}
}
\label{tab:lab_descriptive}
\end{table*}

%% file: tables/lab_gaze_tests.tex
\begin{table*}[!htb]
\centering
\caption{\emph{Post-hoc pairwise comparisons of gaze features across AOIs and rationale conditions.} 
\revtwo{Each cell reports the pairwise $p$ value followed by the effect size in parentheses, $p_{value}$ (effect).} Arrows indicate the direction of the difference (\up{} = first condition $>$ second; \down{} = first $<$ second). Highlighted entries mark $p<.05$ and $p<.01$, respectively. The rationale AOI includes only correct- versus incorrect-rationale trials.}

\label{tab:gaze_tests}
\vspace{-1.0mm}
\footnotesize
\setlength{\tabcolsep}{3pt}
\renewcommand{\arraystretch}{1.160}

\begingroup
% Adjustable column widths
\def\aoiw{1.75cm}
\def\compw{2.25cm}
\def\fixcw{2.05cm}
\def\fixdw{2.05cm}
\def\saccw{2.05cm}
\def\saclenw{2.05cm}
\def\pupow{2.05cm}
\def\pupfw{2.05cm}
\def\ttffw{2.05cm}

\makebox[\linewidth][c]{%
\resizebox{\dimexpr\linewidth-2pt\relax}{!}{%
\begin{tabular}{@{}
>{\centering\arraybackslash}p{\aoiw}
>{\raggedright\arraybackslash}p{\compw}
>{\centering\arraybackslash}p{\fixcw}
>{\centering\arraybackslash}p{\fixdw}
>{\centering\arraybackslash}p{\saccw}
>{\centering\arraybackslash}p{\saclenw}
>{\centering\arraybackslash}p{\pupow}
>{\centering\arraybackslash}p{\pupfw}
>{\centering\arraybackslash}p{\ttffw}
@{}}

\toprule
\textbf{AOI} & \textbf{Pairwise} & \multicolumn{2}{c}{\textbf{Fixation}} & \multicolumn{2}{c}{\textbf{Saccade}} & \multicolumn{2}{c}{\textbf{Pupil Diameter}} & \shortstack{\textbf{TTFF}} \\
\cmidrule(lr){3-4} \cmidrule(lr){5-6} \cmidrule(lr){7-8}
\multicolumn{2}{c}{} & \textbf{Count} & \textbf{Duration} & \textbf{Count} & \textbf{Length} & \textbf{Overall} & \textbf{Fixation} & \multicolumn{1}{c}{} \\

\midrule
\multirow{3}{*}{\textbf{Context}}
& No vs. Corr        & \up{\sigone{.03 (.30)}} & \up{\sigtwo{.01 (.36)}} & \up{.08 (.25)} & \up{.53 (.09)} & \up{.08 (.25)} & \up{.08 (.25)} & \up{.66 (.06)}  \\
& No vs. Incor      & \up{.74 (.05)} & \up{.19 (.18)} & \down{.79 (.04)} & \down{.11 (.22)} & \up{\sigone{.05 (.27)}} & \up{\sigone{.05 (.27)}} & \up{.50 (.09)}  \\
& Corr vs. Incor & \down{\sigone{.05 (.27)}} & \down{.24 (.16)} & \down{\sigone{.04 (.29)}} & \down{\sigone{.03 (.30)}} & \up{.66 (.06)} & \up{.60 (.07)} & \up{.85 (.03)}  \\

\midrule
\multirow{3}{*}{\textbf{Answer}}
& No vs. Corr        & \up{\sigtwo{$<$.001 (1.09)}} & \up{\sigtwo{$<$.001 (.74)}} & \up{\sigtwo{$<$.001 (.59)}} & \up{\sigtwo{$<$.001 (.71)}} & \up{\sigtwo{$<$.001 (.48)}} & \up{\sigtwo{$<$.001 (.45)}} & \down{\sigtwo{$<$.001 (.51)}} \\
& No vs. Incor      & \up{\sigtwo{$<$.001 (.76)}} & \up{\sigtwo{$<$.001 (.61)}} & \up{\sigtwo{$<$.001 (.53)}} & \up{\sigtwo{$<$.001 (.50)}} & \up{\sigtwo{$<$.001 (.56)}} & \up{\sigtwo{$<$.001 (.58)}} & \down{\sigtwo{$<$.001 (.54)}}  \\
& Corr vs. Incor & \down{.09 (.24)} & \down{.29 (.15)} & \down{.30 (.14)} & \down{\sigone{.05 (.27)}} & \up{.45 (.10)} & \up{.89 (.02)} & \down{.70 (.05)}  \\

\midrule
\multirow{1}{*}{\textbf{Rationale}}
& Corr vs. Incor & \down{.23 (.17)} & \down{.11 (.22)} & \down{.19 (.18)} & \down{.66 (.06)} & \down{\sigone{.04 (.28)}} & \down{\sigone{.04 (.29)}} & \up{.32 (.14)}  \\

\midrule
\multirow{3}{*}{\textbf{Response}}
& No vs. Corr        & \up{\sigtwo{$<$.001 (1.15)}} & \up{\sigtwo{$<$.001 (1.05)}} & \up{\sigtwo{$<$.001 (.98)}} & \up{\sigtwo{$<$.001 (.44)}} & \up{\sigone{.02 (.34)}} & \up{\sigone{.04 (.30)}} & \down{.11 (.22)}  \\
& No vs. Incor      & \up{\sigtwo{$<$.001 (1.08)}} & \up{\sigtwo{$<$.001 (1.00)}} & \up{\sigtwo{$<$.001 (1.01)}} & \up{\sigtwo{$<$.001 (.57)}} & \up{\sigone{.02 (.32)}} & \up{\sigone{.02 (.34)}} & \down{\sigtwo{$<$.001 (.50)}}  \\
& Corr vs. Incor & \down{.09 (.23)} & \down{.25 (.16)} & \down{.28 (.15)} & \up{.91 (.02)} & \down{.77 (.04)} & \down{.58 (.08)} & \down{.13 (.21)}  \\

\bottomrule
\end{tabular}
}%
}%
\endgroup

\vspace{-2.0mm}
\end{table*}

%% file: main/5-Study3-methods.tex
\section{Exploratory Predictive Modeling of User States via Gaze Signals}

% =================================================================

Study~2 provided synchronized gaze measures and trial-level self-reported outcomes, enabling a secondary question: whether gaze behavior carries task-specific signal about users' trust-, cognitive-, and decision-related states during LLM-assisted factual verification. This analysis is motivated by the Study~2 gaze results, which showed condition-dependent attention allocation across the context/evidence, answer, rationale, and \revtwo{response/rating} areas, and by prior work using eye-tracking features for task-bounded user-state modeling in HCI and human-AI interaction~\citep{eyetracking_methods,eyetracking_survey,eyetracking_cognitive_1,Ajenaghughrure_modeling,trust_predict_hri,sun_trust}. 
We treat this as a secondary, exploratory predictive analysis of behavioral signals, not as evidence that gaze directly measures trust, cognitive load, or decisions.

% ========================================================================

\subsection{Methods}

\subsubsection{Dataset and Task-Level Sampling}
For each Study~2 factual-verification trial $i$, we created one participant-trial sample. Participants inspected the factual claim and supporting evidence, viewed the LLM answer with either no rationale, a correct rationale, or an incorrect rationale, made a Yes/No decision, and then provided trial-level self-reports. We paired these trial-level outcomes with AOI-level gaze features extracted after preprocessing and with participant-level background covariates from the pre-study questionnaire, yielding one row per participant-trial. After the gaze-data quality exclusion, the predictive modeling dataset contained 636 participant-trial samples from 53 participants, with 12 trials per participant and approximately 212 samples per rationale condition.

\subsubsection{Feature Engineering}
% The predictive feature vector $\mathbf{x}_i = [\mathbf{g}_i, \mathbf{b}_{u(i)}] \in \mathbb{R}^{d}$ combined trial-level gaze features with participant-level background covariates. \revise{The gaze component $\mathbf{g}_i$ comprised seven features---fixation count, fixation duration, saccade count, saccade length, pupil diameter, fixation-level pupil diameter, and time-to-first-fixation---extracted from each available AOI (the claim/evidence, answer, and rationale AOIs, but not the post-response rating area), giving 21 gaze features in the rationale-present conditions and 14 in the no-rationale condition.} The background component $\mathbf{b}_{u(i)}$ comprised demographic indicators and available AI-related questionnaire items from the pre-study questionnaire for participant $u(i)$, repeated across that participant's trials, and entered the models as auxiliary covariates alongside the gaze features. \revise{Condition-specific models used only the AOIs present in that condition, whereas pooled models were fit on samples from all rationale conditions.}

The predictive feature vector $\mathbf{x}_i = [\mathbf{g}_i, \mathbf{b}_{u(i)}] \in \mathbb{R}^{d}$ combined trial-level gaze features with participant-level background covariates. The gaze component $\mathbf{g}_i$ included seven AOI-level features: fixation count, fixation duration, saccade count, saccade length, pupil diameter, fixation-level pupil diameter, and time to first fixation. These features were computed for each available AOI: claim/evidence, answer, rationale when present, and response/rating.

Because the response/rating AOI was included, model performance reflects retrospective trial-level classification rather than pre-decision or real-time user-state prediction.
The background component $\mathbf{b}_{u(i)}$ comprised demographic indicators and available AI-related items from the pre-study questionnaire for participant $u(i)$, repeated across that participant's trials, and entered the models as auxiliary covariates alongside the gaze features. 
% \revise{Condition-specific models used only the AOIs present in that condition, whereas pooled models were fit on samples from all rationale conditions.}
Because the rationale AOI was absent in no-rationale trials, condition-specific models used the AOIs available within each condition: no-rationale models used claim/evidence, answer, and response/rating AOIs, whereas correct- and incorrect-rationale models additionally included the rationale AOI. For pooled models across all rationale conditions, we used only the AOIs shared by all conditions: claim/evidence, answer, and response/rating. This kept the pooled feature schema consistent across trials. 

\subsubsection{Classification Targets}
% We trained separate binary classifiers for five trial-level targets: trust in information, trust in the LLM system, cognitive load, decision accuracy, and decision confidence. Trust in information, trust in the LLM system, cognitive load, and decision confidence were binarized into low/high labels using a median split. Decision accuracy was coded as correct versus incorrect by comparing the participant's Yes/No decision with the ground-truth answer for the factual claim. Because most decisions were correct, decision-accuracy performance should be interpreted against its class base rate rather than against a balanced-chance assumption.
Each prediction task was paired with a set of target variables obtained from the self-reported data:
\vspace{-1.2mm}
\[
\begin{aligned}
\mathbf{s}_i = [&\text{trust-info}_i,\ \text{trust-sys}_i, \text{cognitive-load}_i,\\
&\text{decision-accuracy}_i,\ \text{decision-confidence}_i],
\end{aligned}
\]
and the condition 
$c_i \in \{\text{no rationale},\ \text{correct rationale},\ \text{incorrect rationale}\}$.
This yields the dataset for the predictive modeling task:
\vspace{-1.2mm}
\[
\mathcal{D} = \{(\mathbf{x}_i, c_i, \mathbf{s}_i)\}_{i=1}^{N}.
\]
We trained separate binary classifiers for each target outcome $t$ to learn whether task-level gaze behavior carried predictive information about the corresponding user state:
\vspace{-1.2mm}
\[
f_t(\mathbf{x}_i) \rightarrow y_{i,t},
\]
where $\mathbf{x}_i$ denotes the predictive feature vector for trial $i$. 
Trust in information, trust in LLM system, cognitive load, and decision confidence were binarized into low/high labels using a median split (quantile = 0.5). 
% \revtwo{These median thresholds were computed on the full sample rather than re-estimated within each training fold. Because this uses information from the full label distribution, the classification results should be interpreted as exploratory retrospective analyses; future work should define thresholds within training folds or use externally defined cut points.}
\revtwo{The median thresholds were computed on the full sample rather than within each training fold, creating label-definition leakage because held-out participants contributed to the thresholds used to label their own ratings. We therefore interpret the classification results as exploratory retrospective analyses (see Limitations). 
% Future work should define thresholds within training folds, use external cut points, or model the continuous ratings directly.
}
Decision accuracy was coded as correct versus incorrect by comparing the participant's Yes/No decision with the ground-truth answer. As most decisions were correct, decision-accuracy should be interpreted against its class base rate rather than as a balanced classification task.

\subsubsection{Machine-Learning Classifier Models for the Predictive Tasks}
We evaluated five machine-learning classifiers spanning complementary model families: Logistic Regression (LR) as a linear baseline, Support Vector Machine (SVM) as a kernel-based classifier, Random Forest (RF) as a tree-ensemble classifier, Multilayer Perceptron (MLP) as a neural-network classifier, and Adaptive Boosting (AdaBoost) as a boosting method.

\subsubsection{Model Training and Test Setup}
Models were evaluated with leave-one-subject-out (LOSO) cross-validation~\cite{loso} across participants. 
In each fold, one participant was held out for testing and all remaining participants were used for training. 
% \revtwo{Feature preprocessing, namely standardization of the continuous gaze and background features, was fit only on the training participants in each fold and then applied to the held-out participant.} 
This procedure was used for both pooled and condition-specific predictive analyses, so the reported scores reflect generalization to unseen participants rather than to held-out trials.
% from participants already observed during training. 
\revtwo{The five classifiers were trained with fixed default hyperparameters. 
We deliberately avoided per-fold hyperparameter search to prevent optimistic bias from nested model selection and to keep the comparison across model families transparent. Consequently, the reported scores should be read as conservative estimates under default settings rather than as each model's achievable upper bound.}

For each target, we trained models in four settings. The pooled setting trained one model using trials from all three rationale conditions, using the AOIs shared across conditions. The condition-specific settings trained separate models on the no-rationale, correct-rationale, and incorrect-rationale subsets. These settings were used to examine whether gaze features carried different retrospective signal depending on whether rationales were absent, correct, or incorrect. 
Models were evaluated using gaze-only and gaze-plus-background feature sets by classification accuracy and F1 score~\newcite{f_measure_review}.

\subsubsection{SHAP Analysis}
Lastly, we applied SHAP (SHapley Additive exPlanations)~\citep{xai_shape_raw, xai_shape} as a post-hoc feature-attribution method to the two tree-based models, Random Forest and AdaBoost. \revtwo{To keep the rationale AOI available as a candidate feature, these attributions were computed on the gaze-plus-background models. SHAP results are shown to illustrate feature relevance and are not tied to a single best-performing configuration.}
To keep the interpretation aligned with rationale inspection and response behavior, Fig.~\ref{fig:shap} reports the top AOI-level gaze features identified within these trained models for the trust- and decision-related targets. 
SHAP was used only to visualize which AOI-level gaze features contributed most within these trained models, and was not used for model training, label construction, model selection, or causal inference.

%% file: main/5-Study3-results.tex
\subsection{Results: Predictive Modeling by Classification}

% =============================================================================

\subsubsection{Prediction Performance}

\input{tables/ML_predictions}

% =================================================================

This exploratory predictive modeling tested whether the features described in Section~5.1 carried task-specific predictive signals about user-state labels: cognitive load, trust in information, trust in the LLM system, decision accuracy, and decision confidence. 
Table~\ref{tab:raw_gaze_vs_gazedemo} reports \revtwo{classification accuracy and F1} for both gaze-only and gaze-plus-background feature sets.
Because all labels were derived from post-trial self-reports, these results should be interpreted as retrospective classification, not real-time user-state detection.

\revise{Trust outcomes showed modest performance. With gaze plus background covariates, trust in information reached accuracy up to .67 (F1 .72) in the pooled analysis, and trust in the LLM system reached up to .70 (F1 .70).}
% These values were above chance but remain far below a level that would justify describing gaze as a trust detector; instead, they suggest limited, task-specific signal for trust judgments.}
% Trust outcomes showed modest performance. With gaze plus background covariates, trust in information reached up to .67 accuracy and .72 F1 in the pooled setting, while trust in the LLM system reached up to .70 accuracy and .70 F1. 
\revtwo{These values indicate task-specific signals for predictive modeling that would justify describing gaze as a trust detector, but are still limited.}

% \revise{Cognitive load and decision confidence showed moderate signal. Cognitive load reached accuracy up to .72 (F1 .79) in the pooled model, and decision confidence reached up to .68 (F1 .79). The pooled task context generally carried more stable information than any single rationale condition alone.}
Cognitive load and decision confidence showed somewhat stronger signals. Cognitive load reached an accuracy up to .72 (F1 .79) in the pooled model, and decision confidence reached up to .72 (F1 .82) in the correct condition. 
Across targets, the pooled task context generally carried more stable information than any single rationale condition alone, although this pattern was not uniform across all classifiers and conditions.

% \revise{Decision accuracy produced the highest raw scores, up to .80 (F1 .88) in the pooled setting and .83 (F1 .91) in incorrect-rationale trials. Because most decisions were correct, however, these scores partly reflect the high base rate of correct decisions and should be read as above-baseline retrospective signal rather than as strong evidence of robust outcome prediction.}
Decision accuracy produced the highest raw scores, reaching up to .80 accuracy and .88 F1 in the pooled setting and .84 accuracy and .91 F1 in incorrect-rationale trials. However, because most decisions were correct, these scores should be carefully interpreted relative to the high base rate of correct decisions rather than as strong evidence of robust outcome prediction.

% =================================================================

\begin{figure*}[!ht]
    \centering
    \includegraphics[width=0.999\linewidth]{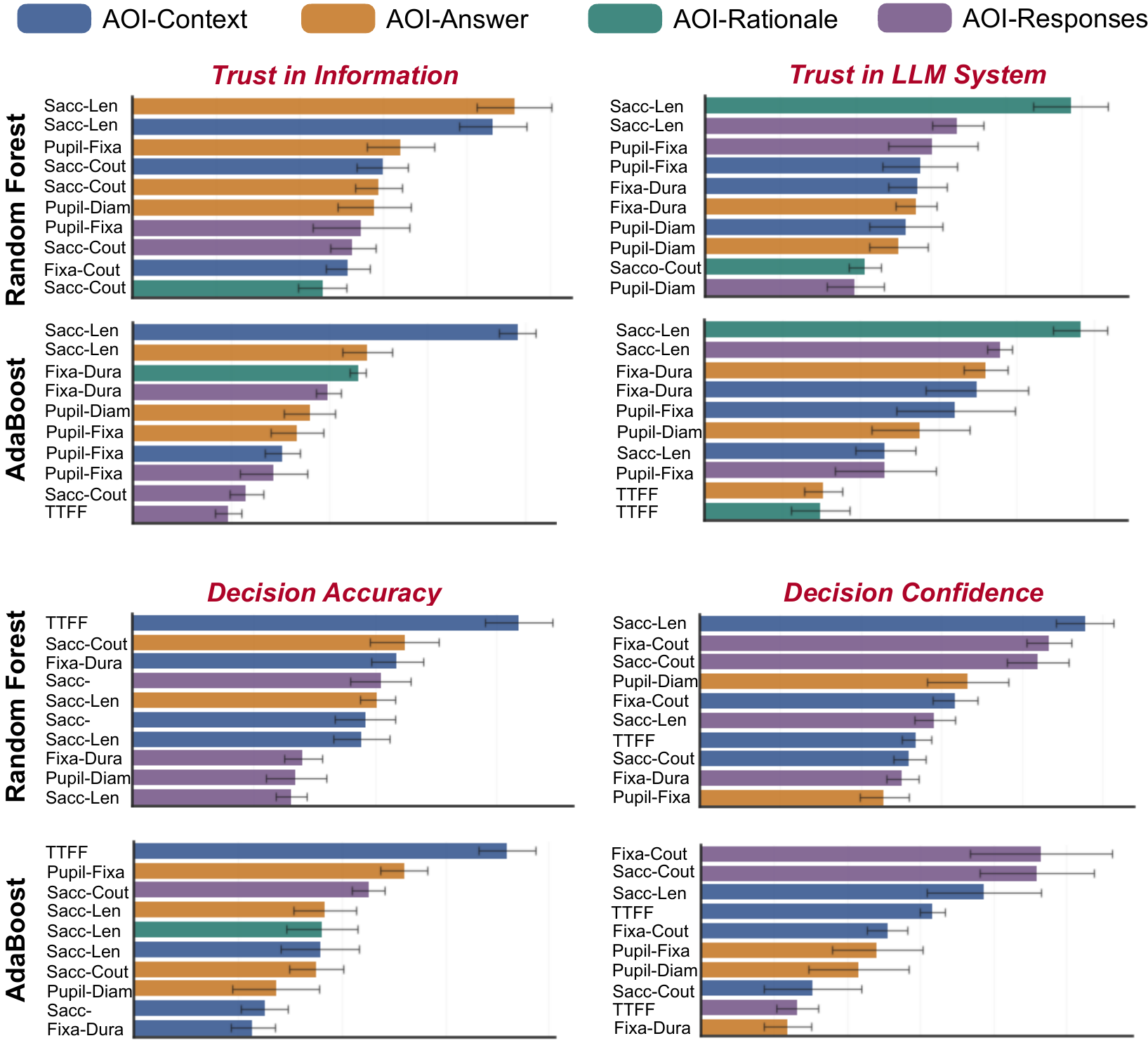}
    \vspace{-1.6mm}
    % \caption{
    % SHAP analysis: top 10 AOI-level gaze features for low/high trust and confidence labels and correct/incorrect decision labels, shown for Random Forest and AdaBoost classifiers. 
    % SHAP values are post-hoc feature attributions within trained models, not causal effects.
    % (``Sacc'' = saccades; ``Fixa'' = fixations; ``Count''/``Cout'' = count; ``Len'' = length; ``Dura'' = duration; ``Diam'' = diameter; ``TTFF'' = time to first fixation.)
    % }
    \caption{\emph{Post-hoc SHAP feature-attribution analysis for exploratory predictive models.} 
    Panels show the top 10 AOI-level gaze features for selected trust- and decision-related targets in the Random Forest and AdaBoost classifiers. These attributions were computed for gaze-only models.
    Bar length indicates the mean absolute SHAP value within the trained model. 
    Colors indicate AOI source: context/evidence, answer, rationale, and response/rating. 
    SHAP values are post-hoc model attributions, not causal effects of real-time user-state detection. 
    Feature abbreviations: Sacc = saccade; Fixa = fixation; Len = length; Dura = duration; Diam = diameter; TTFF = time to first fixation.
    }
    \label{fig:shap}
    \vspace{-0.6mm}
\end{figure*}

% =================================================================

% Taken together, these results suggest that the predictive models captured clearer signal for task-proximal states such as decision accuracy, decision confidence, and cognitive load than for broader evaluative trust judgments in this factual-verification task. 
% The results should therefore be interpreted as evidence of predictive signal in this task, not as evidence that the features directly measure trust or cognition.
% Taken together, these results suggest that the predictive models captured clearer signal for task-proximal states such as decision accuracy, decision confidence, and cognitive load than for broader evaluative trust judgments in this factual-verification task. 
% The results should therefore be interpreted as evidence of predictive signal in this task, not as evidence that the features directly measure trust or cognition.

\revise{Overall, the results indicate that gaze features can support meaningful retrospective classification in this factual-verification context, \revtwo{with several task-proximal outcomes reaching roughly the .70--.84 accuracy range (and F1 up to .91)}. The performance should be interpreted as bounded, context-specific signal rather than robust real-time user-state detection.}

\subsubsection{Post-hoc SHAP Analysis}
The post-hoc SHAP feature-attribution analysis in Fig.~\ref{fig:shap} should be interpreted as exploratory. 
The gaze feature set included AOI-level features from the claim/evidence, answer, rationale when present, and response/rating areas, together with participant-level background variables. 
Prominent attributions were distributed across multiple AOIs rather than concentrated in a single region. 
The figure focuses on the AOI-level gaze features that contributed most within the trained Random Forest and AdaBoost models for the trust- and decision-related targets shown (Fig.~\ref{fig:shap}). 
Across these targets, fixation- and saccade-based features appeared more prominently than pupil-based features. These attributions indicate which features the trained classifiers weighted more heavily. 
They do not establish causal mechanisms or real-time user-state detection or prediction.
% they do not establish causal mechanisms or directly reveal how participants formed trust or confidence judgments.

%% file: tables/ML_predictions.tex
\begin{table*}[!ht]
\centering
\scriptsize
\renewcommand{\arraystretch}{1.10}
\setlength{\tabcolsep}{2pt}
% \caption{Performance of binary classification \textbf{(accuracy/F1)} under \textbf{raw} prediction mode for the \textbf{Gaze-only} and \textbf{Gaze + background} feature sets. LogisticReg = Logistic Regression; SVM = Support Vector Machine; RandomForest = Random Forest; MLP = Multilayer Perceptron. Underlined values indicate the highest value per column; red underlined values indicate the best overall value per user state.}

\caption{\emph{Binary classification performance in the exploratory predictive modeling analysis.} 
Values report accuracy/F1 under LOSO cross-validation for gaze-only and gaze-plus-background feature sets. 
Columns show pooled models across all rationale conditions (Pooled) and condition-specific models for no-rationale, correct-rationale, and incorrect-rationale trials. 
Underlined values indicate the highest score within each column; red underlined values indicate the best observed score for each user-state target. 
Abbreviations: LogisticReg = Logistic Regression; SVM = Support Vector Machine; RandomForest = Random Forest; MLP = Multilayer Perceptron; AdaBoost = Adaptive Boosting.}
\vspace{1.2mm}
\textbf{Panel A. Gaze-only: Trust outcomes}
\vspace{0.8mm}
\begin{tabular*}{\textwidth}{@{\extracolsep{\fill}}l*{8}{c}@{}}
\toprule
\multirow{2}{*}{\textbf{Classifier}} & \multicolumn{4}{c}{\textbf{Trust in Information}} & \multicolumn{4}{c}{\textbf{Trust in LLM System}} \\
\cmidrule(lr){2-5} \cmidrule(lr){6-9}
& \textbf{Pooled} & \textbf{No} & \textbf{Correct} & \textbf{Incorrect} & \textbf{Pooled} & \textbf{No} & \textbf{Correct} & \textbf{Incorrect} \\
\midrule
LogisticReg  & .61/.65 & .55/.58 & .58/.61 & \best{.64/.65} & .56/.58 & .49/.48 & .49/.55 & .56/.59  \\
SVM  & .59/.67 & \underline{.61/.62} & .56/.64 & .58/.60 & .53/.56 & .52/.67 & \best{.62/.71} & .52/.62  \\
RandomForest  & .61/.64 & .59/.60 & .56/.59 & .60/.63 & \underline{.59/.59} & .51/.54 & .50/.59 & \underline{.57/.63}  \\
MLP  & \underline{.61/.65} & .58/.64 & \underline{.62/.65} & .55/.55 & .54/.54 & \underline{.53/.50} & .53/.61 & .54/.61  \\
AdaBoost  & .58/.64 & .58/.59 & .56/.61 & .60/.68 & .58/.61 & .40/.42 & .53/.67 & .54/.64  \\
\bottomrule
\end{tabular*}

\vspace{1.2mm}
\textbf{Panel B. Gaze-only: Cognitive-load and decision outcomes}
\vspace{0.8mm}
\setlength{\tabcolsep}{1.4pt}
\resizebox{\textwidth}{!}{%
\begin{tabular}{l*{12}{c}}
\toprule
\multirow{2}{*}{\textbf{Classifier}} & \multicolumn{4}{c}{\textbf{Cognitive Load}} & \multicolumn{4}{c}{\textbf{Decision Accuracy}} & \multicolumn{4}{c}{\textbf{Decision Confidence}} \\
\cmidrule(lr){2-5} \cmidrule(lr){6-9} \cmidrule(lr){10-13}
& \textbf{Pooled} & \textbf{No} & \textbf{Correct} & \textbf{Incorrect} & \textbf{Pooled} & \textbf{No} & \textbf{Correct} & \textbf{Incorrect} & \textbf{Pooled} & \textbf{No} & \textbf{Correct} & \textbf{Incorrect} \\
\midrule
LogisticReg  & .60/.64 & .63/.65 & .52/.54 & .57/.55 & .57/.69 & .57/.69 & .60/.72 & .54/.67 & .57/.65 & .61/.69 & .61/.69 & .55/.64  \\
SVM  & .61/.76 & .57/.71 & .60/.75 & .47/.40 & \underline{.79/.87} & .74/.86 & \underline{.76/.87} & \best{.82/.89} & \underline{.65/.79} & .67/.80 & .70/.81 & .68/.81  \\
RandomForest  & \best{.65/.74} & \underline{.64/.70} & .57/.68 & \underline{.58/.57} & .77/.86 & \underline{.79/.88} & .75/.85 & .82/.90 & .64/.76 & .65/.77 & .69/.80 & \underline{.70/.81}  \\
MLP  & .60/.71 & .58/.64 & \underline{.60/.74} & .51/.49 & .79/.88 & .78/.88 & .73/.83 & .78/.87 & .64/.76 & .63/.76 & .68/.81 & .68/.81  \\
AdaBoost  & .61/.74 & .59/.66 & .60/.74 & .51/.47 & .79/.88 & .75/.86 & .76/.87 & .82/.90 & .64/.76 & \underline{.68/.81} & \best{.71/.83} & .65/.79  \\
\bottomrule
\end{tabular}%
}

\vspace{2.6mm}
\setlength{\tabcolsep}{2pt}
\textbf{Panel C. Gaze + background: Trust outcomes}
\vspace{0.8mm}
\begin{tabular*}{\textwidth}{@{\extracolsep{\fill}}l*{8}{c}@{}}
\toprule
\multirow{2}{*}{\textbf{Classifier}} & \multicolumn{4}{c}{\textbf{Trust in Information}} & \multicolumn{4}{c}{\textbf{Trust in LLM System}} \\
\cmidrule(lr){2-5} \cmidrule(lr){6-9}
& \textbf{Pooled} & \textbf{No} & \textbf{Correct} & \textbf{Incorrect} & \textbf{Pooled} & \textbf{No} & \textbf{Correct} & \textbf{Incorrect} \\
\midrule
LogisticReg  & .63/.67 & .57/.61 & \underline{.64/.66} & .65/.66 & .67/.67 & \underline{.65/.66} & \underline{.67/.69} & .64/.66  \\
SVM  & .58/.66 & \underline{.61/.62} & .56/.64 & .58/.61 & .53/.56 & .51/.67 & .58/.71 & .52/.61  \\
RandomForest  & .62/.66 & .57/.58 & .61/.65 & .66/.67 & .69/.69 & .63/.65 & .59/.69 & \underline{.69/.73}  \\
MLP  & \best{.67/.72} & .61/.65 & .62/.66 & \underline{.66/.70} & \best{.70/.70} & .57/.48 & .56/.62 & .65/.69  \\
AdaBoost  & .58/.63 & .57/.59 & .52/.55 & .62/.68 & .65/.66 & .56/.57 & .58/.67 & .64/.67  \\
\bottomrule
\end{tabular*}

\vspace{1.2mm}
\textbf{Panel D. Gaze + background: Cognitive-load and decision outcomes}
\vspace{0.8mm}
\setlength{\tabcolsep}{1.4pt}
\resizebox{\textwidth}{!}{%
\begin{tabular}{l*{12}{c}}
\toprule
\multirow{2}{*}{\textbf{Classifier}} & \multicolumn{4}{c}{\textbf{Cognitive Load}} & \multicolumn{4}{c}{\textbf{Decision Accuracy}} & \multicolumn{4}{c}{\textbf{Decision Confidence}} \\
\cmidrule(lr){2-5} \cmidrule(lr){6-9} \cmidrule(lr){10-13}
& \textbf{Pooled} & \textbf{No} & \textbf{Correct} & \textbf{Incorrect} & \textbf{Pooled} & \textbf{No} & \textbf{Correct} & \textbf{Incorrect} & \textbf{Pooled} & \textbf{No} & \textbf{Correct} & \textbf{Incorrect} \\
\midrule
LogisticReg  & .65/.69 & .62/.65 & .58/.63 & \underline{.68/.69} & .61/.72 & .67/.77 & .62/.73 & .67/.78 & .60/.67 & .64/.71 & .61/.71 & .61/.70  \\
SVM  & .61/.76 & .57/.71 & .60/.75 & .48/.41 & .79/.88 & \underline{.73/.84} & \underline{.77/.87} & \best{.84/.91} & .65/.78 & .67/.80 & .70/.81 & \underline{.68/.81}  \\
RandomForest  & \best{.72/.79} & \underline{.64/.71} & .57/.67 & .59/.61 & .78/.88 & .75/.86 & .76/.85 & .83/.91 & .65/.76 & .62/.73 & \best{.72/.82} & .67/.79  \\
MLP  & .67/.76 & .62/.70 & .54/.65 & .54/.44 & \underline{.80/.88} & .71/.82 & .70/.81 & .78/.87 & .65/.78 & .64/.76 & .68/.81 & .60/.73  \\
AdaBoost  & .66/.75 & .60/.65 & \underline{.60/.74} & .52/.50 & .79/.88 & .75/.86 & .74/.85 & .82/.90 & \underline{.68/.79} & \underline{.67/.80} & .71/.83 & .67/.79  \\
\bottomrule
\end{tabular}%
}
\vspace{-1.0mm}
\label{tab:raw_gaze_vs_gazedemo}
\vspace{-0.0mm}
\end{table*}

%% file: main/6-Discussion.tex
\section{Discussion}

This work shows that user-facing LLM rationales are not simply additional explanations. Once shown to users, rationales become interface cues that influence \revtwo{trust perception and decision-making behaviors}. 
Across two studies, the strongest effects came less from how rationales were disclosed than from their epistemic quality: whether reasoning was correct, whether it expressed certainty, and whether users could inspect it against the answer. We discuss these findings through the lens of \revtwo{auditable trust calibration} and derive \revtwo{design considerations for LLM-powered decision-support systems}.

% ==================================================================================

\subsection{LLM Rationales Are Double-Edged Trust-Relevant Cues}

% \revise{In Study~1, no reliable presentation-format or interaction effects emerged, whereas rationale correctness increased trust in the information, trust in the LLM system, and decision confidence. Certainty framing primarily shifted trust in the LLM system.} In Study~2, rationale correctness again affected trust in the information and trust in the LLM system, while decision accuracy and decision confidence did not reliably differ across rationale conditions.
In Study~1, presentation format showed no reliable effect in the planned analyses, whereas rationale correctness increased trust in the information, trust in the LLM system, decision confidence, and advice adoption. Certainty framing also shifted trust-, confidence-, and adoption-related judgments. In Study~2, rationale correctness again affected trust in the information and trust in the LLM system, while decision accuracy and decision confidence did not reliably differ across rationale conditions.

% This distinction is important for trust calibration. Prior work on automation and AI-assisted decision-making argues that the goal is not simply to increase trust, but to support appropriate reliance: users should rely when the system is reliable and withhold reliance when it is not~\citep{measure_trust_automation,Wischnewski}. Our findings connect this logic to user-facing LLM rationales. Rationales can change how much users trust or follow AI advice, but these changes do not necessarily translate into better objective performance. \revise{Indeed, Study~2 did not show reliable differences in decision accuracy or decision confidence across rationale conditions, and the decision-confidence effect in Study~1 was correspondingly small, indicating that rationale correctness shaped trust judgments more consistently than decision confidence.} This is consistent with prior work showing that explanations and confidence cues can increase reliance or subjective confidence without improving decision quality~\citep{confidence_effect_in_decision_1,explanation_ai_overreliance,llm_uncertainty,measure_confidence,kim2024uncertainty,si2024verify,kim2025reliance}.
This distinction is important for trust calibration. Prior work on automation and AI-assisted decision-making argues that the goal is not simply to increase trust, but to support appropriate reliance: users should rely when the system is reliable and withhold reliance when it is not~\citep{measure_trust_automation,Wischnewski}. Our findings connect this logic to user-facing LLM rationales. Rationales can change how much users trust or follow AI advice, but these changes do not necessarily translate into better objective performance. Study~2 did not show reliable differences in decision accuracy or decision confidence across rationale conditions, and the decision-confidence effect of rationale correctness in Study~1 was small. This suggests that rationale quality shaped trust judgments more consistently than it improved decision outcomes. The pattern is consistent with prior work showing that explanations and confidence cues can increase reliance or subjective confidence without necessarily improving decision quality~\citep{confidence_effect_in_decision_1,explanation_ai_overreliance,llm_uncertainty,measure_confidence,kim2024uncertainty,si2024verify,kim2025reliance}.

For LLM-powered systems, this creates a double-edged effect. A correct rationale may help users see why an answer is warranted and support more appropriate reliance. However, a fluent, incorrect, or overconfident rationale may also make reasoning appear more reliable than it is. This risk is especially important because LLM rationales are natural-language explanations: they can look like careful reasoning even when they are incomplete, inconsistent, or unfaithful to the answer they appear to justify~\citep{reasoning_incorrect_1,reasoning_incorrect_2,inconsistency_llm_reasoning}. In this sense, rationales are not transparency benefits by default. They are trust-relevant cues with calibration potential, whose value depends on whether they help users judge when reliance is warranted.

% The findings also suggest that rationale design should not be evaluated only by whether users like the explanation or whether trust increases. A rationale that raises trust in advice supported by flawed reasoning is not a successful explanation. A rationale that lowers trust because it reveals a real flaw may be useful, even if it reduces acceptance. The more relevant question is whether the rationale helps users align trust, confidence, and advice adoption with the quality of the answer and the reasoning. \revise{This is the design goal of auditable trust calibration: rationales should help users decide when to rely, when to question, and when to check the evidence.} \revtwo{We note, however, that because the LLM's answer was always correct in both studies, the current data speak to how rationale quality modulates trust and adoption when the advice is reliable. They do not yet show whether rationale quality helps users reject unreliable advice, which is the other half of the calibration problem.}
The findings also suggest that rationale design should not be evaluated only by whether users like the explanation or whether trust increases. A rationale that increases trust without helping users inspect the relationship among answer, evidence, uncertainty, and reasoning is not necessarily successful. Conversely, a rationale that lowers trust because it exposes flawed reasoning may be useful, even if it reduces acceptance. The more relevant question is whether the rationale helps users align trust, confidence, and advice adoption with the quality of the answer and reasoning. This is the design goal of auditable trust calibration: rationales should help users decide when to rely, when to question, and when to verify. 
A critical implication of this design is that audit-like inspection should not be equated with successful calibration. 
Because the LLM's answer was always correct in both studies, however, the current data speak to how rationale quality influences trust, adoption, and inspection under reliable advice. They do not yet show whether rationale quality helps users reject unreliable advice, which remains the other half of the calibration problem.

% Because the LLM's final answer was always correct, lower trust or lower advice adoption after an incorrect rationale cannot be interpreted straightforwardly as improved calibration. It may reflect useful sensitivity to flawed reasoning, but it may also reflect unwarranted doubt toward a correct answer. In this sense, two interpretations are empirically indistinguishable in the present design: flawed rationales may prompt users to audit the evidence, but they may also lead users to discount reliable advice. We therefore interpret the gaze findings as evidence of rationale-driven inspection and trust modulation, not as evidence that rationales improved calibrated reliance.

% ==================================================================================

% \subsection{Rationales Can Be Audited, Not Just Read: Gaze Behavior and User-State Modeling}
\subsection{LLM Rationales Can Be Audited, Not Just Read}
Study~2 adds process-level evidence to the findings observed in Study~1. Self-reported trust and decision-related outcomes show how users evaluated information, but they do not reveal how users inspected the answer, rationale, and supporting evidence while forming those judgments. The gaze data help address this gap. No-rationale trials led to greater inspection of the answer and \revtwo{response/rating} areas than rationale-present trials, although these contrasts should be interpreted cautiously because the no-rationale interface also differed in layout and available text. More diagnostically, incorrect rationales were associated with more context-oriented scanning than correct rationales, and larger fixation-level pupil diameter in the rationale AOI.

% These patterns suggest that rationales may have changed users' verification strategies. The simplest interpretation is not that users blindly accepted rationales.
% Rather, users appeared to adjust their attention depending on whether the reasoning looked reliable or flawed. 
% \revise{Compared with correct rationales, incorrect rationales appeared to redirect attention toward the evidence and to require more cognitive effort, while the rationale AOI itself did not show reliable correctness differences in fixation frequency or duration.} 
% This interpretation is consistent with prior eye-tracking work showing that fixation, saccade, and pupil measures can provide evidence of attention allocation and cognitive effort, while not directly revealing trust or belief~\citep{eyetracking_methods,eyetracking_cognitive_1,Cacioppo}.
These patterns suggest that user-facing rationales can become objects of audit rather than merely text to be read. 
The clearest correctness-specific gaze pattern was not that participants looked longer at incorrect rationales themselves, but that incorrect rationales were associated with more attention to the context area and larger pupil responses while viewing the rationale. 
This is consistent with additional checking or effortful processing, while remaining short of proving a specific mental state. Prior eye-tracking work similarly shows that fixation, saccade, and pupil measures can provide process evidence about attention allocation and cognitive effort, but should not be treated as direct measures of trust or belief~\citep{eyetracking_methods,eyetracking_cognitive_1,Cacioppo}.
However, audit-like inspection is not necessarily beneficial by itself. Under reliable answers, increased checking after a flawed rationale may either support detection of problems in the reasoning trace or create unnecessary doubt about a correct answer. 
Thus, the gaze findings identify a process relevant to auditable calibration, but they do not establish that calibration was causally achieved.

% This process evidence supports that user-facing rationales can become audit objects. Participants were not only exposed to rationale text; they used the information around the rationale. 
% They \revise{allocated attention across} the answer, rationale, evidence, and rating areas in ways that \revise{were consistent with} different forms of inspection. 
% This matters because prior XAI work argues that explanations help most when users can compare them with the underlying evidence at reasonable cost~\citep{explanation_ai_overreliance,impact_xai_decision_making,users_perceive_cot,bucinca2021}. 
% Our gaze findings \revise{are consistent with this logic at the behavioral level}. A rationale is more useful when users can check what it claims, where it draws support from, and whether it conflicts with the supporting context.
This process evidence strengthens auditability framing. 
Participants were not only exposed to rationale text, their gaze patterns also indicate that they inspected information around the rationale, in ways consistent with checking whether the reasoning was supported. This aligns with prior XAI work arguing that explanations are most useful when users can compare them with underlying evidence at reasonable cost~\citep{explanation_ai_overreliance,impact_xai_decision_making,users_perceive_cot,bucinca2021}. 
A user-facing rationale is therefore more useful when it makes claims that can be checked, connects those claims to evidence, and exposes conflicts or uncertainty rather than merely presenting a fluent rationale.

% The predictive modeling analysis adds a narrower point. Gaze features \revise{showed task-specific predictive signal that varied by target. Decision accuracy produced the highest scores, although this partly reflects the high base rate of correct decisions; cognitive load and decision confidence were moderate, and the trust labels were the most modest, with trust in information the hardest target}. This pattern makes sense: AOI-level gaze features are closer to trial-level inspection and response behavior than to broader evaluations of system competence. \revise{However, because the labels are post-hoc, these results still require further validation, such as pre-decision feature windows, before they could inform real-time adaptation.} 
% Prior sensing work similarly shows that gaze and physiological data can support user-state modeling, but only for specific targets and under specific task conditions~\citep{Ajenaghughrure_modeling,trust_predict_hri,sun_trust}.
The predictive modeling analysis adds a narrower point. Gaze features showed task-specific retrospective signal, but the signal varied by target. Decision accuracy produced the highest scores, cognitive load and decision confidence showed moderate signal, and trust-related labels were more modest. This pattern is plausible because AOI-level gaze features are closer to trial-level inspection and response behavior than to broader evaluations of system competence. It echoes the prior sensing work similarly suggesting that gaze and physiological signals can support user-state modeling for specific targets, tasks, and interaction contexts~\citep{Ajenaghughrure_modeling,trust_predict_hri,sun_trust}.

% More important, this result should be interpreted cautiously. The predictive models do not show that gaze measures trust. The SHAP analysis only describes which features helped the trained classifiers separate labels in this dataset; it does not establish psychological causation. 
% A longer fixation on a rationale may reflect careful checking or confusion. 
% More movement between rationale and context may reflect verification, uncertainty, or inconsistency monitoring. 
% The safer conclusion is that gaze can provide task-bounded behavioral signal about checking, effort, and decision-related states when interpreted together with rationale quality and certainty cues.
% % It should not be treated as a general trust detector or as a standalone trigger for adaptive intervention.
The SHAP analysis should be read with the same caution. It identifies which gaze features were useful within trained classifiers, but it does not establish psychological causation. Longer fixation, larger pupil diameter, or more movement between AOIs may reflect checking, uncertainty, confusion, or task demands depending on context. 
The safer conclusion is that gaze can provide bounded, task-specific behavioral signal about inspection, effort, and decision-related behaviors when interpreted together with rationale quality and interface context, rather than being treated as a general trust detector.

% ==================================================================================

\subsection{Design Considerations for LLMs with Rationales}

Across the studies, the design implication is not simply to show more reasoning. User-facing rationales should be designed with auditable trust calibration as a goal: they should help users decide whether an answer is warranted, what needs verification, and how much weight to place on the model's expressed certainty. This framing builds on prior work showing that explanations are useful when they support appropriate reliance, not merely when they increase perceived understanding, trust, or reliance~\citep{impact_xai_decision_making,explanation_ai_overreliance,responsible_ai,ux_responsible}. Four design considerations follow.

% \textbf{Prioritize rationale reliability over presentation polish.}
% \revise{Study~1 showed that users were influenced more by rationale correctness than by whether rationales were shown instantly, delayed, or on demand. Certainty cues also shifted system-level trust, but the strongest survey-study effects remained tied to whether the rationale was correct.} This does not mean presentation format is irrelevant in all settings. It means that disclosure mechanics alone are unlikely to address the core trust-related problem. Designers should first ask whether the rationale is logically consistent, evidence-grounded, and appropriately uncertain. A polished reveal interaction cannot compensate for a rationale that is fluent but wrong, unsupported, or inconsistent with the answer.
% This is consistent with prior work showing that explanations can change reliance without improving objective performance, and that chain-of-thought style rationales may still be plausible but unfaithful to the answer they appear to justify~\citep{confidence_effect_in_decision_1,explanation_ai_overreliance,reasoning_incorrect_1,reasoning_incorrect_2,inconsistency_llm_reasoning}.
\textbf{Prioritize rationale quality over presentation format.}
Study~1 found little reliable evidence that presentation format alone changed outcomes, whereas rationale correctness and certainty framing influenced trust- and decision-related judgments. This does not mean that presentation format is irrelevant in all settings. Rather, it suggests that reveal mechanics alone cannot solve the core trust problem. Designers should first ask whether the rationale is consistent, evidence-grounded, and \revtwo{appropriately calibrated in its uncertainty}.
A polished presentation cannot compensate for reasoning that is fluent but wrong or inconsistent with the answer. This aligns with prior work showing that explanations can change reliance without improving decision quality, and that chain-of-thought-style rationales can appear plausible while remaining incomplete, unfaithful, or inconsistent~\citep{confidence_effect_in_decision_1,explanation_ai_overreliance,reasoning_incorrect_1,reasoning_incorrect_2,inconsistency_llm_reasoning}.

% \textbf{Treat rationales as audit traces, not persuasive narratives.}
% The qualitative reports and Study~2 gaze patterns both suggest that participants used rationales to check the model's reasoning rather than to outsource judgment wholesale. 
% Incorrect rationales redirected attention toward context and were associated with more effortful verification. 
% This aligns with work showing that explanation use depends on whether users can inspect the explanation at reasonable cost and connect it to the decision evidence~\citep{impact_xai_decision_making,explanation_ai_overreliance,users_perceive_cot}. 
% Rationales should therefore support inspection: reasoning should be presented in stepwise units, assumptions should be visible, and each step should be easy to compare with the supporting evidence. 
% The goal is not to make rationales sound more fluent, but to make them easier to verify.
\textbf{Treat rationales as audit traces, not persuasive narratives.}
The qualitative responses and Study~2 gaze patterns both suggest that users can use rationales as material for checking, rather than simply as reasons to outsource judgment. Incorrect rationales were associated with more context-oriented scanning and larger pupil responses in the rationale AOI, consistent with additional evidence inspection and processing effort. Rationales should therefore be structured to support audit: reasoning steps should be separable, assumptions should be visible, and claims should be easy to compare with the supporting evidence. The goal is not to make rationales sound more fluent or human-like, but to make them easier to verify.

% \textbf{Calibrate certainty cues to rationale quality.}
% Certainty framing had clear effects on trust, decision confidence, and advice adoption in Study~1. This makes certainty language a high-impact interface signal rather than decorative wording. If certainty cues are not aligned with actual rationale quality, they may produce miscalibrated trust. Prior work on uncertainty communication similarly shows that verbal confidence cues can shape user trust and metacognitive calibration independently of objective correctness~\citep{llm_uncertainty,measure_confidence,confidence_effect_in_decision_1}. Interfaces should therefore avoid strong confidence language when reasoning quality is uncertain, and should communicate uncertainty in ways that are specific and actionable. For example, uncertainty should indicate what is uncertain---the evidence, the inference, or the final answer---rather than simply lowering users' trust through a vague ``I am unsure'' cue.

\textbf{Calibrate certainty cues to rationale quality.}
\revtwo{Certainty framing was a consequential interface cue} in Study~1, affecting trust- and decision-related judgments. This makes certainty framing more than decorative wording. If certainty cues are not aligned with the quality of the answer, rationale, or evidence, they may contribute to miscalibrated reliance~\citep{llm_uncertainty,measure_confidence,kim2024uncertainty}. Interfaces should therefore avoid strong confidence language when reasoning quality is uncertain. Uncertainty should also be specific and actionable: users should be told whether uncertainty comes from weak evidence, ambiguous inference, possible answer error, or unreliable rationale construction.

\textbf{Use behavioral sensing as contextual support.}
The predictive analysis suggests that gaze features can carry retrospective signal in this factual-verification context, especially for task-proximal outcomes. 
% However, the signal was still target-dependent and not sufficient for real-time user-state detection. 
Behavioral sensing may be useful for support functions such as surfacing evidence links, prompting verification when users appear to engage in effortful checking, or offering additional scaffolding around rationales. 
Any gaze-informed adaptation should be transparent, opt-in, privacy-preserving, and oriented toward user autonomy rather than manipulating reliance~\citep{Ajenaghughrure_modeling,trust_predict_hri,sun_trust,Cacioppo}.

% Taken together, these design considerations do not suggest that interfaces should always show more reasoning. Rather, they suggest showing reasoning when it remains warranted, inspectable, and low-cost to verify. In other words, rationale design should optimize for auditable calibration rather than maximal disclosure or rhetorical persuasiveness~\citep{Chen_Jing,responsible_ai}.

Taken together, these design considerations argue against maximal disclosure or rhetorical persuasiveness as design goals~\citep{Chen_Jing,responsible_ai}. Rationales should be shown when they are warranted, inspectable, and low-cost to verify. 
The goal of rationale design should be auditable calibration: helping users connect answer, evidence, uncertainty, and reasoning so they can decide when to rely, when to question, and when to check.

% ==================================================================================

\subsection{Limitations and Future Work}

% This work has several limitations we need to acknowledge and that define the scope of the findings and suggest directions for future work.
This work has several limitations that define the scope of the findings and motivate future work.

\paragraph{\textbf{Study task and stimulus scope}}
For experimental control, we used short binary factual-verification claims with pre-generated, author-curated rationales. We manipulated a focused set of rationale properties: presentation format, correctness, and certainty framing in Study~1, and rationale presence and correctness in Study~2. The findings therefore apply most directly to controlled factual-verification tasks and may not generalize to open-ended, higher-stakes, or multi-turn LLM interactions, or to other rationale properties such as length, writing style, source provenance, evidence linking, and visual layout. Study~1 also used a small item pool of \revtwo{six counterbalanced items, one per condition}, so its condition effects are partly confounded with item difficulty and should not be interpreted as item-generalized estimates. 
% Study~2 used a demographically narrow laboratory sample and relatively few trials per condition, which further limits generalizability and the precision of item-level estimates. 
\revtwo{Because Study~1 used a gender-balanced online sample, whereas Study~2 used a younger, predominantly female laboratory sample, the convergent rationale pattern across the two studies should be read as conceptual replication under different populations rather than as a single accumulating effect.}

\paragraph{\textbf{LLM answer scope}}
Because the LLM's answer was always correct, the studies cannot test whether users appropriately reject incorrect AI advice. \revtwo{This design choice allowed us to isolate the effect of rationale quality while holding answer reliability constant, so that changes in trust, decisions, and gaze behavior could be attributed more directly to the reasoning trace rather than to answer correctness.} 
This creates an interpretive ambiguity within the present data: when incorrect rationales lowered trust or adoption, this may indicate sensitivity to flawed reasoning, but it may also indicate unwarranted doubt toward a correct answer. Incorrect rationales in our studies therefore represent flawed reasoning attached to reliable advice, not unreliable advice itself. The findings do not establish successful trust calibration across both reliable and unreliable advice. 
Future work should cross answer correctness with rationale correctness to distinguish productive auditing from miscalibrated rejection of reliable advice.

\paragraph{\textbf{Interface and gaze-measurement scope}}
In the on-demand condition, we did not log whether participants revealed the rationale or how long they viewed it. \revtwo{The three presentation formats were therefore not fully equivalent in enforced rationale visibility, which may have diluted format effect}, so the absence of presentation-format effects should not be over-interpreted. 
% \revtwo{In Study~2, the no-rationale condition by design contains no rationale region, so it differs from the rationale-present conditions in layout and on-screen text; we therefore read no-rationale versus rationale-present gaze contrasts as descriptive and base the correctness interpretation on the correct- versus incorrect-rationale comparisons, which share an identical layout.} More generally, gaze measures should be interpreted as process indicators of visual attention and effort, not as direct measures of trust, belief, or reasoning quality.
\revtwo{In Study~2, the no-rationale condition contained no rationale region by design. We therefore interpret no-rationale versus rationale-present gaze contrasts as descriptive condition-level comparisons that reflect differences in both information layout and available text. The correctness interpretation is based primarily on correct- versus incorrect-rationale comparisons, which share the same AOI structure and interface layout.} 
More generally, gaze measures should be interpreted as process indicators of visual attention and effort, not as direct measures of trust or reasoning quality.

\paragraph{\textbf{Exploratory modeling scope}}
The predictive analysis was secondary and exploratory. 
Because labels were derived from post-trial self-reports, the results should be interpreted as a limited retrospective, task-specific classification signal rather than real-time trust detection. 
\revtwo{In addition, the binary labels were defined using a single median threshold computed on the full sample rather than re-estimated within each cross-validation fold. Although this threshold depends only on the label distribution, it constitutes a mild form of label-definition leakage that we did not eliminate.
Future work should compute thresholds within each training fold or use externally defined cut points.} 
The models also require stronger validation against appropriate baselines, uncertainty estimates, and external datasets before they can support deployable user-state sensing and prediction. 
Future work should evaluate pre-decision feature windows, majority-class and permutation baselines, and larger cross-task validation sets to determine whether gaze features generalize beyond the present interface and task.

Overall, across these boundaries, future work should use larger and more diverse participant samples and item pools, test naturally generated rationales in interactive multi-turn systems, and examine more realistic decision-support settings where both answer correctness and rationale quality vary.

%% file: main/7-Conclusion.tex
\section{Conclusion}

This work investigates how user-facing LLM rationales influence user trust and gaze patterns during LLM-assisted factual verification. 
% Across controlled studies, correct rationales generally increased trust relative to incorrect rationales.
\revise{In Study~1, rationale correctness had the clearest survey effects, increasing trust in the information, trust in the LLM system, and decision confidence, while certainty framing also shifted trust and decision-related outcomes.} 
In Study~2, incorrect rationales also lowered system trust relative to no rationale, while the no-rationale comparison was weaker for trust in information. 
% \revise{In Study~2, however, these trust-related differences did not translate into reliable gains in objective decision accuracy or decision confidence.} 
Additionally, gaze features \revise{showed post-hoc, task-specific predictive signal that was clearest for decision accuracy and modest for trust-related labels}, motivating future work on user-aware systems that adapt rationale and certainty presentation after careful validation of behavioral signals. 
Overall, our findings caution against ``always showing more explanation'' and support selective, audit-oriented rationale presentation that helps users inspect whether a rationale is warranted, evidence-linked, and appropriately certain.